\documentclass[aps,pra,twocolumn]{revtex4}%
\usepackage{amsmath}
\usepackage{amsfonts}
\usepackage{amssymb}
\usepackage{graphicx}%
\providecommand{\U}[1]{\protect\rule{.1in}{.1in}}

\begin{document}
\title{Tunneling electro-conductance of atomic Bose condensates}
\author{V. M. Akulin$^{1}$, Yu.\ E.\ Lozzovik$^{2}$, I. E. Mazets$^{3,4}$, A. G.
Rudavets$^{5}$, and A. Sarfati$^{1}$}
\affiliation{$^{1}$Laboratoire Aim\'{e} Cotton, CNRS, Campus d'Orsay, 91405, Orsay, France}
\affiliation{$^{2}$Insitute of Spectroscopy, 142190, Moscow Reg. Troitsk, Russia}
\affiliation{$^{3}$A.F. Ioffe Physics-Technical Institute, St. Petersburg 194021, Russia}
\affiliation{$^{4}$Atominstitut der \"{O}sterreichischen Universit\"{a}ten, TU Wien,
A--1020 Vienna, Austria}
\affiliation{$^{5}$Moscow Institute of Physics and Technology, 141700 Dolgoprudny, Russia}

\pacs{}

\begin{abstract}
We consider interaction of an electron with a Bose condensate of atoms having
electron affinity. Though states of the electron attached to atoms form a
continuous band, tunneling through this band is strongly suppressed by quantum
fluctuations of the condensate density. We adapt standard field theory methods
originally developed for description of a particle propagating trough a
disordered potential and present an exactly soluble analytical model of the
process. In contrast with the standard description, we take into account
inelastic processes associated with quantum transitions in the condensate.
Possibilities of the experimental observation of the phenomenon are discussed.

\end{abstract}
\date[Date text ]{\today }
\maketitle

\section{Why electro-conductance of atomic condensates?\ }

Of course a simple curiosity first of all. But in order to provide a deeper
motivation, we review the milestones in the history of Bose condensation from
the first days till present time, trying to convince readers that this
question appears as a natural step in the development of this fundamental
field of research.

In 1924 S.\ N.\ Bose \cite{Bose} has calculated partition of the radiation
field (Termodynamische Wahrscheinlichkeit f\"{u}r Strahlungsfeld) employing
combinatorial formula for indistinguishable elements. Two weeks
later\ Einstein \cite{Einstein} has generalized this approach to the case of
material particles forming an ideal one-atom gases and has noticed that it
yields a decrease of the number of particles at a given velocity as compared
to the prediction of the Maxwell law. Six months later, Einstein had stated
directly that Bose distribution below a certain temperature is incompatible
with the conservation of the number of particle, unless one assumes that some
fraction of particles is condensed in the state with zero momentum. These
classical results were purely thermodynamical being based only on the notion
of the size of phase volume per one quantum state. The De Broglie wave aspects
of the condensation have not been discussed.

The fundamental aspects of the field theory required \ for description of
atomic Bose-Einstein condensates, such as collective quantum
states\cite{Landau}, weakly interacting quasiparticles \cite{Bogolyubov}, and
the mean-field description \cite{Landau1, Pitaevskii, Gross}, have been
developed in the middle of XX-th century in the context of superfluidity. But
only at the end of the century it become possible to address the condensation
of one-atom gases experimentally \cite{Exp, Exp1, Exp2}.

For the radiation field, the situation turned out to be the opposite, --
discovery of masers and lasers took place in years 50-th that is much earlier
than understanding of a deep analogy between lasing from one hand side and
phase transitions\cite{Scully} from the other hand side, including the Bose
condensation of photons \cite{Oraevskii} as a particular case of phase
transitions. By now it became clear that the coherent laser radiation results
from condensation of photons in the same mode of an optical resonator with an
active media inside. But as in the case of Bose condensation of gases, the
first explanations of lasing were also almost thermodynamical being based on
the rate equation and relations between the Einstein's kinetic coefficients
for spontaneous and induced radiations. The quantum state aspects of the
radiation field have not been addressed.

However, the rate equations alone cannot account for the coherence properties
of the radiation, since they completely ignore the important phase relations
among the quantum states of the field.\ The consistent quantum-mechanical
consideration of lasing shows that the resulting quantum state of the
electromagnetic field mode approaches a coherent Glauber state\cite{Glauber}
with minimum energy-phase uncertainty, which goes to the classical
limit\cite{Sudarshan} when the number of quanta in the mode increases.
Therefore within the kinetic approach, the statement that the induced photons
are all coherent, whatever it means, has been taken as a sort of axiom,
whereas the classical description\cite{Lamb} of the electromagnetic field
above the lasing threshold was considered as satisfactory for all practical
purposes. Quantum nature of the radiation\cite{Mandel} was only addressed in
the context of the radiation noises and the photon counting
statistics\cite{Lax}.

Though the theory required for the description of atomic Bose condensates has
been developed earlier, from the standpoint of quantum-classical
correspondence it is constructed in the same way as the laser theory. Based on
the Ginzburg-Landau approach to phase transitions\cite{Landau1, Ginzburg}, it
relies on the collective atomic amplitude satisfying Gross-Pitaevskii
equation. This amplitude\ being a quantum object by itself, still should be
considered as a classical field in the context of quantum variables of
individual atoms, - following Dirac\cite{Dirac},\ the corresponding quantum
field operators in the second quantization representation are replaced by a
large classical field plus relatively small quantum fluctuations. By making
use of the Bogolyubov transformation, these fluctuations are partially taken
into account and result in a Vlasov-like dispersion relation between the
quasiparticles energies and the momenta, while the rest of the contribution
accounting for quantum noises is usually ignored.

The collective amplitude approach is absolutely adequate for the present
experimental situation \cite{FS}, and traditionally it is employed for
predictions and explanations of many fascinating results observed
experimentally, such as condensate interference\cite{interBEC, interBEC1,
interBEC2}, vortices\cite{vortex, vortex1}, sound\cite{sound}, excited
states\cite{excitations} hydrodynamic condensate motion\cite{hydrodyn,
hydrodyn1}, including that in the presence of long-range interatomic
interactions.\ Moreover, this amplitude satisfying the nonlinear
Schr\"{o}dinger equation, invites to extend classical results of the Condensed
Matter physics on the case of atomic condensates. The solid-state phenomena,
such as quantum wires\cite{wires}, Anderson localization\cite{localization,
localization1, localization2, localization3, localization4, localization5},
Josephson contacts\cite{JC}, nanotransistors\cite{atomtonics, atomtonics1,
atomtonics2}, metal-dielectric phase transition\cite{Fazio, Fazio1, Fazio2,
Fazio3, Fazio4}, band structures\cite{else, else1}, etc. have already been
discussed in the context of condensed atoms. The non-linearity of the
Schr\"{o}dinger equation enreaches the variety of possible effects that can be
observed. Still, most of these phenomena remain classical from the viewpoint
of the second quantization, since they do not address the essentially quantum
field-theoretical aspects of the atomic Bose condensates, such as quantum fluctuations.\ 

Thermodynamical fluctuations have been considered at the very early stage of
the development of Bose condensates theory\cite{Furst, London}, including
their De Brogle wave aspects\cite{Galanin}. Also the kinetic aspects of Bose
condensations have already been discussed\cite{kin, kin1, kin2, Kocharovskii2}%
. Moreover, following natural logics of the development of Bose condensate
theory, one inevitably arrives at a point where consideration of the
essentially quantum fluctuations becomes crucial. One of the examples is the
condensation dynamics. According to the Liouville theorem, the condensation
associated with a shrinking of the occupied phase volume cannot occur in a
closed system, -- one needs to invoke a dissipation and hence fluctuations
associated with the dissipation in virtue of the fluctuation-dissipation
theorem. In the context of lasing this problem has been intensively studied in
years 60-th\cite{ScullyLamb} and by now is well presented in
textbooks\cite{Relaxation}. Recently a consistent description of the Bose
condensation dynamics in the spirit of the laser theory has been
developed\cite{Kocharovskii, Kocharovskii1}.

In this work we consider another example, where quantum field fluctuations
play the dominating role. It concerns an electron tunneling through the Bose
condensate at the energy close to the electron affinity $E_{a}$ of the
condensed atoms. We show that the electro-conductance associated with this
process is indeed strongly influenced by an essentially quantum quantity - the
size of atomic density fluctuations. The quantum localization
mechanism\cite{anderson} underlying this strong influence is well understood
in the theory of disordered media. In our consideration we rely on two tools
that have emerged from this theory: the concept of tunneling transparency
\cite{LIFSHITS PASTUR} and the $\sigma$-model technique\cite{Efetov}, although
both these tools being relevant to the case of an electron moving in a
classical disordered potential have to be modified in order to include the
case of random potentials created by the essentially quantum fluctuations.
Indeed, according to the Born-Oppenheimer principle, the electron sees not the
quantum averages of the atomic positions given by the mean field but their
instantaneous values. The latters, when being in generic positions, form a
disordered media for the tunneling electron. The averaging over the atomic
quantum state should therefore be performed only for the final result of
consideration of the electronic part of problem -- for the probability of
tunneling through the disordered media\cite{Inelastic}.

Drawing parallels with the laser physics, one can say in a very broad sense,
that the discreetness of atoms plays a role similar to the shot noise of a
laser beam photons registered by a radiation detector. However it turns out
that the naive analog of antibunching in photon counting, namely the binary
correlation of the atomic positions in condensate\cite{Mazets}, is not the
parameter that governs the tunneling conductance.\ The latter depends on the
size of collective quantum density fluctuations. Focussing at the simplest
case, which can be exactly solved analytically, we consider only the situation
where the ensemble of $\mathcal{N}$ condensed atoms is in a coherent state.
From the second quantization point of view, this state is the direct analog of
the Glauber state for the laser field with the average number $\mathcal{N}$ of
photons. An analog of spontaneous photons, the non-condenced fraction of the
gas, is ignored. For soilving the electronic part of the problem, we also
employ the simplest version of $\sigma$-model, which corresponds to Gaussian
unitary ensembles. We believe that in spite of such restrictive assumptions,
this consideration still remains general enough to reveal the main qualitative
effects associated with electro-conductance of Bose condensates.

We note that several questions about charged condensates have already been
addressed. Transformations induced by the presence of atomic ions has been
considered in Ref.\cite{lukin}. The question of a highly efficient charge
exchange in cold gases closely relevant to the tunneling conductance of
condensates was the subject of Ref.\cite{chexch}. Eigenstates and dynamics of
an electrons trapped by an atomic Bose condensate has been considered in
Ref.\cite{DYKHNE} in the framework of the mean field approximation and the
scattering length model of the atom-electron interaction. All these papers
were focus on the case where the charge is completely localized inside the
condensate. However since the polarization energy of atoms usually exceed
considerably the condensate chemical potential, by putting a charge inside a
cold gas one either destroies completely the condensed phase or, in the most
optimistic scenario, triggers a long-lasting process of relaxation to a new
metastable condensed configuration. In contrast, the tunneling conductance
does not imply localization of the charge inside the condensate, and therefore
it appears as the most gentle way of addressing the essentially quantum
properties of condensates by means of an electrical interaction.

The paper is organized as follows. In Sect\ref{I} we present a heuristic
picture of the tunneling trough a condensate, which not pretending to be
rigorous gives an intuitive idea about the process. In Sect.\ref{II} we
introduce the main formal tools required for consideration, the Hamiltonian,
the initial conditions, and describe a representation most adequate to the
case of the electron energy close to $E_{a}$, based on the quantum fields of
atoms and a negative ion. In Sect.\ref{III} we consider the process with the
help of the mean-field approximation and determine the typical time and energy
domains where the tunneling can occur. In Sect.\ref{IV} we turn to the role of
the quantum density fluctuations and show that they crucially modify the
mean-field picture. Trying to make this section as simple for reading as
possible, we have moved to Appendix\ref{1} the most important but rather
technical part of the calculations based on the $\sigma$-model. In
Sect.\ref{V} \ we continue by considering the regime required for the
observation of the tunneling and estimate the atomic density, the typical
tunneling time, the typical current, and typical losses of the condensed atoms
during the tunneling along with there energy dependence. It turns out that in
order to clearly observe the phenomenon one needs to increase the densities of
the atomic condensate by a few orders of magnitude with respect to ones
experimentally available at the moment. We also discuss some requirements that
the leads materials should satisfy in order to ensure efficient transport of
electrons to the condensate.\ In Sect.\ref{VI} we conclude by discussing the
main results obtained and the feasibility of experimental observation of the phenomenon.

\section{A heuristic description\label{I}}

\begin{figure}
[h]
\begin{center}
\includegraphics[
height=2.1594in,
width=2.7518in
]%
{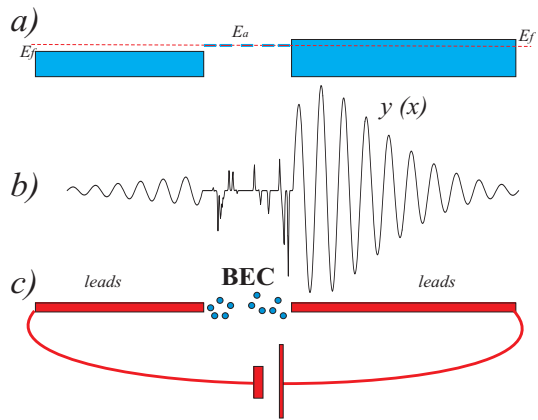}%
\caption{Tunneling conductance of Bose condensate. A wave packet of the
tunneling electron is much larger as compared to the size of the condensate,
such that the electron mainly remains within the leads and no charge present
on atoms during the process. The Fermi energies of the leads \ $E_{f}$ are
close to the electron affinity $E_{a}$ of the atoms. (a) Energy diagram of the
leads and states corresponding to various locations of the negative ion. (b)
Wave packet of the tunneling electron. (c) General view of the setting.}%
\label{CoCoWaveFunction}%
\end{center}
\end{figure}

The tunneling regime implies that the wave function of the tunneling electron
forms a wave packet of a size much larger then the size of the condensate,
such that no considerable charge present at the atoms during the elementary
act of tunneling. This is illustrated in Fig.\ref{CoCoWaveFunction}.
The electron wave packet comes from the right to the point $r_{a}$ , where the
right lead approaches the condensate, and mainly scatters back. Still a small
part of the wave packet amplitude tunnels to the left point $r_{b}$, where the
second lead approaches the condensate. The tunneling occur through a number of
the negative ion states corresponding to the electron location near the
positions $r_{n}$ of the condensate atoms. Each of the ionic states is formed
as a result of the attraction of the electron by a short-range affinity
potentials, which we chose in the form $E_{a}\delta(r-r_{n})$. Penetration of
the electron through the gas is a coherent process resulting from the
interference of all possible tunneling trajectories.

We incorporate the well-known approach to the quantum tunneling
electro-conductence trough a disordered media resulting from the interference
of many possible tunneling trajectories into the quantum consideration of the
condensate. Briefly, the idea of the description is following. As the first
step, for a fixed realization of random potential, one writes a formal
expression
\[
G_{A}(\left\{  U(r)\right\}  ,\eta,r_{a},r_{b})G_{R}(\left\{  U(r)\right\}
,\eta,r_{b},r_{a})
\]
in terms of advanced and retarded Green's functions for the probability of
tunneling of an electron at energy $\eta$ from an initial point $r_{a}$ to a
final point $r_{b}$ through the domain of random potential $U(r)$, which
includes interference of all possible tunneling trajectories. As the second
step, one takes an average over all possible realization of the random
potential given by the functional integral
\begin{equation}
\int G_{A}(U,\eta,r_{a},r_{b})G_{R}(U,\eta,r_{b},r_{a})W(U)\mathcal{D}U(r)
\label{Tuu}%
\end{equation}
with a functional wight $W(U)$ that gives the probability of realization of
the potential $U(r)$. This allows\ one to get rid of unimportant details
related to particular realizations and retain only the universal contribution
of the trajectory interference. After such averaging, the problem can usually
be traced analytically. In one- and two-dimensional setting, the tunneling is
considerably affected by the phenomenon of strong and weak quantum
localization, respectively, that result from the interference of self
intersecting trajectories.

Tunneling of an electron through the Bose condensate must be very similar to
the tunneling through a disordered potential
\begin{equation}
U(r)=\sum_{n=1}^{\mathcal{N}}E_{a}\delta(r-r_{n}) \label{PotEn}%
\end{equation}
with a random distribution of all $\mathcal{N}$ atomic positions $r_{n}$,
although this process should have an important difference, -- the average has
to be performed not over an ensemble of the all possible realization of the
random potential, but over the quantum distribution of the condensate atoms.
If after the tunneling, the condensate remains in the initial quantum state,
the analogy is complete, whereas the quantum atomic density distribution
$\rho(\left\{  r_{n}\right\}  )$ plays the role of weight function :%
\[
\int G_{A}(\left\{  r_{n}\right\}  ,\eta,r_{a},r_{b})G_{R}(\left\{
r_{n}\right\}  ,\eta,r_{b},r_{a})\rho(\left\{  r_{n}\right\}  )d^{\mathcal{N}%
}r.
\]
The situation gets much richer when one considers the tunneling accompanied by
quantum transitions in the condensate. In this case, one has to develop an
approach, which takes into account all possible final states of the
condensate, including the outcomes that correspond to a partial or complete
condensate destruction. In the framework of the Born-Oppenheimer separation of
the electronic and atomic motions, the tunneling probability associated with
the condensate transition from the initial state with the condensate wave
function $\psi_{in}(\left\{  r_{n}\right\}  )$ to the final state $\psi
_{fn}(\left\{  r_{n}\right\}  )$ reads%
\begin{align}
&  \int\psi_{fn}(\left\{  r_{n}\right\}  )G_{A}(\left\{  r_{n}\right\}
,\eta,r_{a},r_{b})\psi_{in}^{\ast}(\left\{  r_{n}\right\}  )d^{\mathcal{N}%
}r\label{ProbTun}\\
&  \times\int\psi_{fn}^{\ast}(\left\{  r_{n}\right\}  )G_{R}(\left\{
r_{n}\right\}  ,\eta,r_{b},r_{a})\psi_{in}(\left\{  r_{n}\right\}
)d^{\mathcal{N}}r,\nonumber
\end{align}
and the summation over all possible final states yields the overall tunneling probability.

\begin{figure}
[h]
\begin{center}
\includegraphics[
height=2.1646in,
width=2.8798in
]%
{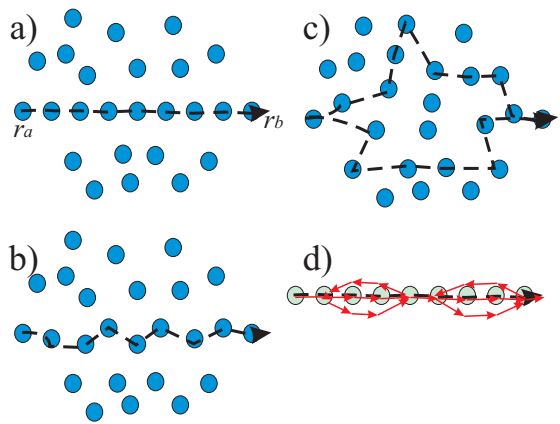}%
\caption{Tunneling trails for various atomic configurations resulting from
quantum density fluctuations in the initial quantum state of BEC. Interference
of all the trajectories that go (forward and back) along the same trail yields
the net trail conduction. (a) A rare highly conducting fluctuation : atoms are
allined in an equidistant one dimensional chain. (b) A more probable
fluctuation : the tunneling chain is not equidistant. The conduction is thus
suppressed by Anderson localization. (c) The most likely situation: many
trails, each of which is weekly conducting, contribute to the overall
conductance. (d) A trajectory can go forward and back along the trail.}%
\label{InterfeRTraj}%
\end{center}
\end{figure}

Let us consider a heuristic picture of such processes comparing several
situations shown in Fig.\ref{InterfeRTraj}.%
Let us assume that the condensate is initially in a pure quantum state given
by the wave function $\psi_{in}(\left\{  r_{n}\right\}  )$. Distribution of
the atomic density in this state results from the superposition of the
probability amplitudes of various atomic configurations, including the rare
ones that correspond to high probability of the electron tunneling. In other
words, for the condensate in a pure quantum state, the highly conducting
configurations may appear as a result of unlikely quantum fluctuations of the
atomic density. In Fig.\ref{InterfeRTraj} a) we show one of such unlikely
configurations, -- some of the condensate atoms form a one dimensional
equidistant chain, which serves as a conducting trail connecting the points
$r_{a}$ and $r_{b}$. Tunneling through such a chain results from the
constructive interference of all electron trajectories going along the trail
(one of them is shown in Fig.\ref{InterfeRTraj} d)), and the overall tunneling
time approximately equals to the time of tunneling between neighboring atoms
multiplied by the number of atoms in the chain.

Evidently, the probability of an exactly equidistant chain is vanishing small,
but it increases exponentially when we consider configurations where the
atomic positions may deviate from the equidistant ones, as shown in
Fig.\ref{InterfeRTraj} b). However, such deviations destroy the constructive
interference of the tunneling trajectories, and the chain conductance gets
exponentially suppressed by the effect of Anderson localization. When the
allowed non-equidistance in the chain increases, the effect of exponentially
growing probability is compensated by the exponential decrease of its
tunneling transparency, such that the net contributions of different
configurations becomes comparable. To answer the question "which sort of the
trail configurations gives the main contribution?", one has to take into
account more subtile dependences, and in particular the dependence on the
spacial dimensionality of the problem. Moreover, one cannot exclude \textit{a
priori} the dominating contribution of numerous interfering "diffusive"
trails, shown in Fig.\ref{InterfeRTraj} c).

Dominating trajectories of different type result in different degree of the
condensate destruction. In fact, a rapid tunneling implies existence of an
almost equidistant highly conducting chain. (Fig.\ref{InterfeRTraj} a). The
"which way" question is thus answered, at least partially, and the condensate
atoms are somehow localized on the trail connecting the points $r_{a}$ and
$r_{b}$. According the standard prescriptions of Quantum Mechanics, the
initial state of the condensate therefore gets reduced. The fact of a rapid
passage of the electron through the tunneling interval can be considered as a
sort of measurement applied to the condensate. Note that this measurement is
indirect, and therefore it is not associated with a Hermitian operator of a
physical quantity. The relevant quantity is the matrix element of the electron
evolution operator
\[
\int\psi_{fn}^{\ast}(\left\{  r_{n}\right\}  )G_{R}(\left\{  r_{n}\right\}
,t,r_{b},r_{a})\psi_{in}(\left\{  r_{n}\right\}  )d^{\mathcal{N}}r,
\]
given by the inverse Fourier transform of the Green's function $\int
\mathrm{d}\eta\exp[\mathrm{i}\eta t]G_{R}(\left\{  r_{n}\right\}  ,\eta
,r_{b},r_{a})$. As the result of measurement, the atoms localized at the
tunneling trail got extracted from the condensate, and the elementary act of
tunneling thus implies evaporation of all the chain atoms. In contrast, in the
opposite limit when the dominating contribution comes from interference of
numerous long trajectories (Fig.\ref{InterfeRTraj} c), the "which way"
question remains without answer, and the correspondent condensate state
reduction is much less pronounced. It can result in evaporation of just a few
atoms per the tunneling event. In order to understand which scenario takes
place in reality, one has to solve the problem analytically.

\section{Hamiltonian and the initial states\label{II}}

Let us start the analytic consideration of the condensate conductance by
reviewing the relevant fundamentals of the quantum many-body theory of Bose
liquids. The description rely on the quantum field theory, where the atomic
variables are given in terms of the bosonic second quantization operators
acting on the vacuum state $\left\vert 0\right\rangle $ : the creation field
operator $\hat{\Phi}^{+}(r)$ and the annihilation field operator $\hat{\Phi
}(r)$. The atomic density at the point $r$ corresponds to the mean value of
the operator $\widehat{n}(r)=\hat{\Phi}^{+}(r)$ $\hat{\Phi}(r)$. With the help
of Fourier transformation one introduces the corresponding field operators
$\hat{\Phi}^{+}(p)$, $\hat{\Phi}(p)$ and $\widehat{n}(p)=\hat{\Phi}^{+}(p)$
$\hat{\Phi}(p)$\ in the momentum representation. The Hamiltonian of the
ensemble of cold atoms%
\[
\hat{H}_{0}=\int\hat{\Phi}^{+}(p)\epsilon_{p}\hat{\Phi}(p)\mathrm{d}%
^{d}p+\gamma\int\hat{\Phi}^{+}(r)\hat{\Phi}(r)\hat{\Phi}^{+}(r)\hat{\Phi
}(r)\mathrm{d}^{d}r
\]
involves the single particle kinetic energy given by the dispersion law
$\epsilon_{p}$ and the local binary interaction characterized by a coupling
constant $\gamma=\frac{4\pi\hbar^{2}a_{s}}{M}$ given in terms of the
scattering length $a_{s}$ Here $M$ is the atomic mass, and $\hbar$ is the
Planck constant, which we set to unity hereafter unless the contrary is said explicitly.

In the limit of large number of particles $\mathcal{N}\rightarrow\infty$ one
can employ the semiclassical approximation for the field variables assuming
the condensate in a coherent field state \cite{Lifshits}
\[
\left\vert A(r)\right\rangle =\exp\left\{  \mathrm{i}\int\left[  A(r)\hat
{\Phi}^{+}(r)+A^{\ast}(r)\hat{\Phi}(r)\right]  \mathrm{d}^{d}r\right\}
\left\vert 0\right\rangle ,
\]
which results from the application of the field displacement operator%
\begin{equation}
\widehat{U}_{d}\left(  A(r)\right)  =\mathrm{e}^{\mathrm{i}\int\left[
A(r)\hat{\Phi}^{+}(r)+A^{\ast}(r)\hat{\Phi}(r)\right]  \mathrm{d}^{d}r}.
\label{displacement}%
\end{equation}
to the vacuum state $\left\vert 0\right\rangle $. The variational principle
$\frac{\delta}{\delta A^{\ast}(r)}\left\langle A(r)\right\vert \hat{H}%
_{0}-E\left\vert A(r)\right\rangle =0$ then yields the Gross-Pitaevskii
equation%
\[
EA(r)=\epsilon(\widehat{p})A(r)+\gamma\left\vert A(r)\right\vert ^{2}A(r)
\]
for \ the displacement field eigen functions $A(r)$. Here $\epsilon
(\widehat{p})$ is the dispersion law $\epsilon_{p}$ where the momentum is
replaced by the momentum operator.\ In the presence of an external field the
potential energy term has also to be included.

The lowest energy solution of the Gross-Pitaevskii equation gives the
fundamental state of the condensate. However, in zero order Born-Oppenheimer
approximation the excitations do not directly contribute to the condensate
conductance as long as the main parts of atoms remain in the fundamental
state. They may manifest themselves indirectly, via the mean density
$\left\langle n(r)\right\rangle $ and the mean fluctuation of the particle
number $\delta n=\left(  \left\langle n^{2}(r)\right\rangle -\left\langle
n(r)\right\rangle ^{2}\right)  ^{1/2}$. Therefore here, we ignore the
excitations and consider the technically simplest case of a coherent initial
state of the condensate%
\begin{equation}
\left\vert in\right\rangle =\exp\left\{  \mathrm{i}\int\left[  A\hat{\Phi}%
^{+}(r)+A^{\ast}\hat{\Phi}(r)\right]  \mathrm{d}^{d}r\right\}  \left\vert
0\right\rangle , \label{Initial}%
\end{equation}
with a uniform mean density $n=AA^{\ast}$ and Gaussian local fluctuations
$\delta n\propto n^{1/2}$ . Moreover, remaining within the Born-Oppenheimer
approximation, we will ignore dynamics of the Bose condensate during the
electron tunneling process, and hence the initial state of the condensate
Eq.(\ref{Initial}) is the only ingredient required for the further consideration.

Note that the initial state Eq.(\ref{Initial}) results from the displacement
of the condensate vacuum state $\left\vert 0\right\rangle $ in the phase space
of the second quantization, which is given by the displacement operator
$\widehat{U}_{d}\equiv\widehat{U}_{d}\left(  A(r)=\mathrm{const}\right)  $ of
Eq.(\ref{displacement}) with a spacially uniform displacement field $A(r)$,
whereas the vacuum state corresponds to zero average number of atoms and only
allows for the quantum fluctuations of the atomic density.

Let us now turn to the description of the tunneling electron. The fermionic
field creation $\widehat{\psi}^{+}(r)$ and$\ $annihilation$\ \widehat{\psi
}(r)$ operators and their counterparts $\widehat{\psi}^{+}(p)$ and$\ \widehat
{\psi}(p)$ in the momentum representation together with the potential energy
Eq.(\ref{PotEn}) yield the electronic part of the Hamiltonian
\begin{align}
\hat{H}_{e}  &  =\int\widehat{\psi}^{+}(p)\frac{p^{2}}{2m}\widehat{\psi
}(p)\mathrm{d}^{d}p\label{ElectronHamiltonian}\\
&  +E_{a}\int\hat{\Phi}^{+}(r^{\prime})\hat{\Phi}(r^{\prime})\delta\left(
r-r^{\prime}\right)  \widehat{\psi}^{+}(r)\widehat{\psi}(r)\mathrm{d}%
^{d}r^{\prime}\mathrm{d}^{d}r\nonumber
\end{align}
which also includes interaction with the atoms.\ Here $m$ is the electron
mass. However, this form of the Hamiltonian is not very convenient for the
description of the slow electron tunneling at the energies close to $E_{a}$.
In fact, the electron states relevant to the tunneling conductance correspond
to the wave functions that are strongly localized near the positions of the
condensate atoms, and therefore the mean values of both kinetic and the
potential parts of Eq.(\ref{ElectronHamiltonian}) are large, and just their
difference is of the order of the inverse tunneling time. In high enrgy
physics this situation corresponds to the well-known positronium problem.
Acting by the analogy, we introduce the second quantization\ operators of the
negative ions%
\begin{align}
\hat{\Psi}(r)  &  =\hat{\Phi}(r)\int\varphi(r-r^{\prime})\widehat{\psi
}(r^{\prime})\mathrm{d}^{d}r^{\prime}\nonumber\\
\hat{\Psi}^{+}(r)  &  =\hat{\Phi}^{+}(r)\int\varphi(r-r^{\prime})\widehat
{\psi}^{+}(r)\mathrm{d}^{d}r^{\prime} \label{Formfactor}%
\end{align}
with the help of the wave function $\varphi(r-r^{\prime})$ of the electron
attached to an isolated atom. In this representation the Hamiltonian
Eq.(\ref{ElectronHamiltonian}) shifted by the electron affinity $E_{a}$ takes
the form
\begin{equation}
\hat{H}_{e}=\int_{\mathcal{V}}\mathrm{d}^{d}r\int_{\mathcal{V}}\mathrm{d}%
^{d}r_{1}\hat{\Psi}^{+}(r)\hat{\Phi}(r)V(r-r_{1})\hat{\Psi}(r_{1})\hat{\Phi
}^{+}(r_{1}) \label{HAMILTONIAN}%
\end{equation}
where $\mathcal{V}$ is the system volume. The interaction operator $V(r)$ has
the explicit physical meaning of the electron tunneling probability amplitude
between two condensate atoms at a distance $r$. This amplitude is given by the
electron affinity multiplied by the overlap of electron wave functions
$\varphi$ centered at different atoms.

It is convenient to work the momentum representation, where the field
operators are related to that in the coordinate representation by the Fourier
transformations%
\begin{align*}
\hat{\Psi}(r)  &  =\int\hat{\Psi}(p)\mathrm{e}^{\mathrm{i}pr}\frac
{\mathrm{d}^{d}p}{\left(  2\pi\right)  ^{d}}\\
\hat{\Psi}(p)  &  =\int\hat{\Psi}(r)\mathrm{e}^{-\mathrm{i}pr}\mathrm{d}^{d}r.
\end{align*}
In this representation the interaction operator%
\[
V(p)=\int\mathrm{d}^{d}r\mathrm{e}^{-\mathrm{i}pr}V(r)
\]
it has an explicit form%
\begin{equation}
V(p)=\frac{\kappa^{2}\Lambda}{p^{2}+\kappa^{2}}, \label{Interaction}%
\end{equation}
identical for all one, two, and three dimensional cases. Here $\kappa
=\sqrt{2mE_{a}}$ is a parameter in the asymptotic expression $\exp\{-\kappa
r\}$of the atomic electron wave function corresponding to the electron
affinity $E_{a}$, and $\Lambda\sim2E_{a}/\kappa^{d}$ is a parameter related to
the shape of the bound electron wave function $\varphi(r-r_{n})$ near the
neutral atom. Note that in the coordinate representation the kernel of the
interaction operator reads%
\begin{equation}
V(r)=\frac{1}{\left(  2\pi\right)  ^{d}}\int\mathrm{e}^{\mathrm{i}%
pr}V(p)\mathrm{d}^{d}p=\left\{
\begin{array}
[c]{c}%
\Lambda\kappa\mathrm{e}^{-\kappa\left\vert r\right\vert }%
~~~~~~\mathrm{~for~1D}\\
\frac{\Lambda\kappa^{2}}{\pi}K_{0}(\kappa\left\vert r\right\vert
)~\mathrm{for~2D}\\
\frac{\Lambda\kappa^{2}}{\pi\left\vert r\right\vert }\mathrm{e}^{-\kappa
\left\vert r\right\vert }~~~~~\mathrm{for~3D}%
\end{array}
\right.  , \label{INTERACTION}%
\end{equation}
where $K_{0}$ is the Bessel function.

Numerical value of the parameters $\Lambda$ and $\kappa$ entering the
interaction Eq.(\ref{Interaction}) can be found either by direct calculation
in the framework of the model of $\delta$-potential, or by comparison of the
calculated charge exchange cross section with the experimentally observed
values. The last option better takes into account the realistic potential that
binds the electron to the neutral atom. For $Cs$ \ atoms with the electron
affinity $E_{a}=0.39eV$ in the dimensional units one finds%
\begin{equation}
\kappa=\frac{\sqrt{2mE_{a}}}{\hslash}=2.26\ 10^{9}\left[  m^{-1}\right]  .
\end{equation}
The coupling constant $\Lambda$ can be found from the value of the velocity
dependent resonant charge transfer cross section%

\begin{equation}
\sigma(v)=\int\limits_{0}^{\infty}2\pi r\sin^{2}\left\{  \int\limits_{-\infty
}^{\infty}\frac{\Lambda\kappa^{2}}{\pi\hbar\sqrt{r^{2}+v^{2}t^{2}}}%
\mathrm{e}^{-\kappa\sqrt{r^{2}+v^{2}t^{2}}}\mathrm{d}t\right\}  \mathrm{d}r,
\end{equation}
which in the limit of high velocity $v$ yields%
\begin{equation}
\sigma(v)=\frac{4\Lambda^{2}\kappa^{2}}{\pi v^{2}\hbar^{2}}.
\end{equation}
By comparing this expression with the experimental data of Ref.\cite{Bydin}
$\sigma=500\pi\mathring{A}^{2}$ at the kinetic energy $1KeV$ ($v=3.8\quad
10^{4}\frac{m}{s}$) \ one finds%
\[
\Lambda=3.9\quad10^{-47}\left[  \frac{m^{5}}{s^{2}}kg\right]  ,
\]
and hence for the 3D case%
\[
\frac{\Lambda\kappa^{2}}{\pi\left\vert r\right\vert }\mathrm{e}^{-\kappa
\left\vert r\right\vert }\simeq0.9\left[  eV\right]  \frac{\exp\left(
-2.26\ 10^{9}r\left[  m\right]  \right)  }{2.26\ 10^{9}r\left[  m\right]  }.
\]
Note that the $\delta$-potential model yields smaller tunneling amplitudes,
which we will employ later for the pessimistic estimations.

Hamiltonian Eq.(\ref{HAMILTONIAN}) with the interaction Eq.(\ref{INTERACTION})
and the initial state of the condensate Eq.(\ref{Initial}) have to be
complemented by the initial $\hat{\Psi}^{+}(r_{a})\left\vert 0\right\rangle
_{i}$ and the final $\left\langle 0\right\vert _{i}\hat{\Psi}(r_{b})$ states
of the ion that correspond to the elementary act of the conduction, where
$\left\vert 0\right\rangle _{i}$ is the ionic vacuum state. Here we consider
only the case of a single electron tunneling, leaving the problem of the
Cooper pairs tunneling for future studies. Therefore the the ion density
operator - $\hat{\Psi}(r)\hat{\Psi}^{+}(r)$ and the atom density operator
$\hat{\Phi}(r)\hat{\Phi}^{+}(r)$ given in terms of the field operators satisfy
the normalization conditions%
\begin{align}
\left\langle 0\right\vert _{i}\int_{\mathcal{V}}\hat{\Psi}(r)\hat{\Psi}%
^{+}(r)\mathrm{d}^{d}r\left\vert 0\right\rangle _{i}  &  =1\label{1.1a}\\
\left\langle 0\right\vert _{i}\int\hat{\Psi}(p)\hat{\Psi}^{+}(p)\frac
{\mathrm{d}^{d}p}{\left(  2\pi\right)  ^{d}}\left\vert 0\right\rangle _{i}  &
=1\\
\int_{\mathcal{V}}\left\langle A(r)\right\vert \hat{\Phi}(r)\hat{\Phi}%
^{+}(r)\left\vert A(r)\right\rangle \mathrm{d}^{d}r  &  =\mathcal{N}%
\label{1.1b}\\
\int\left\langle A(p)\right\vert \hat{\Phi}(p)\hat{\Phi}^{+}(p)\left\vert
A(p)\right\rangle \frac{\mathrm{d}^{d}p}{\left(  2\pi\right)  ^{d}}  &
=\mathcal{N}%
\end{align}

\section{Mean field approximation for the tunneling dynamics\label{III}}

We now consider the probability of the electron tunneling through the Bose
condensate in the framework of the simplest mean field model. The results of
these calculations are strictly speaking incorrect, since they get strongly
modified when, in the next Section, we take into account quantum fluctuations
of the condensate. Still the model is useful for identification of typical
tunneling times and typical structures of the tunneling probability dependence
on the electron energy. Though in the chosen regime of large electron wave
packets (Fig.\ref{CoCoWaveFunction}) the electron is never completely
localized in the condensate, we will employ the dynamic formulation of the
problem where the electron is initially localized at an atom near the point
$r_{a}$ and, as a result of sequential tunneling from one atom to another,
reaches an atom near the point $r_{b}$ . This representation technically is
much more convenient for certain calculations, while the results for the
regime of large wave packets can be obtained by proper time convolutions. In
other words, when considering the tunneling electron we neglect its Coulombic
effect on the translational degrees of freedom of the condensate atoms.

The initial state of the system
\begin{equation}
\left\vert in\right\rangle =\hat{\Phi}(r_{a})\hat{\Psi}^{+}(r_{a}%
)\mathrm{e}^{\mathrm{i}\int\left[  A\hat{\Phi}^{+}(r)+A^{\ast}\hat{\Phi
}(r)\right]  \mathrm{d}^{d}r}\left\vert 0\right\rangle \left\vert
0\right\rangle _{i} \label{1.2}%
\end{equation}
reveals the fact that the electron by entering the condensate in the state
Eq.(\ref{Initial}) annihilates an atom in the point $r_{a}$ and creates an ion
at it's place. By the analogy, the final state of the system just before the
electron leaves the condensate reads
\begin{equation}
\left\langle fn\right\vert =\left\langle 0\right\vert _{i}\left\langle
0\right\vert \mathrm{e}^{-\mathrm{i}\int\left[  A\hat{\Phi}^{+}(r)+A^{\ast
}\hat{\Phi}(r)\right]  \mathrm{d}^{d}r}\hat{\Phi}^{+}(r_{b})\hat{\Psi}(r_{b}).
\label{1.2a}%
\end{equation}
The time dependent tunneling probability amplitude given by the retarding
Green's function $G_{R}(t,r_{a},r_{b})=\left\langle fn\right\vert \exp\left\{
-\mathrm{i}t\hat{H}\right\}  \left\vert in\right\rangle $ has the explicit
form%
\begin{align}
&  G_{R}(t,r_{a},r_{b})=\left\langle 0\right\vert _{i}\left\langle
0\right\vert \mathrm{e}^{-\mathrm{i}\int\left(  A\hat{\Phi}^{+}(r)+A^{\ast
}\hat{\Phi}(r)\right)  \mathrm{d}^{d}r}\hat{\Phi}^{+}(r_{b})\hat{\Psi}%
(r_{b})\nonumber\\
&  \exp\left\{  -\mathrm{i}t\int\mathrm{d}^{d}r\mathrm{d}^{d}r_{1}\hat{\Psi
}^{+}(r)\hat{\Phi}(r)V(r-r_{1})\hat{\Psi}(r_{1})\hat{\Phi}^{+}(r_{1})\right\}
\nonumber\\
&  \hat{\Phi}(r_{a})\hat{\Psi}^{+}(r_{a})\mathrm{e}^{\mathrm{i}\int\left[
A\hat{\Phi}^{+}(r)+A^{\ast}\hat{\Phi}(r)\right]  \mathrm{d}^{d}r}\left\vert
0\right\rangle \left\vert 0\right\rangle _{i}. \label{1.3}%
\end{align}

Now we employ the representation suggested by the unitary displacement
operator $\widehat{U}_{d}$ of Eq.(\ref{displacement}), which with the
allowance for the bosonic commutation relations $\left[  \hat{\Phi}%
(r),\hat{\Phi}^{+}(r^{\prime})\right]  =\delta(r-r^{\prime})$, results in the
transformations
\begin{align}
\hat{\Phi}^{+}(r^{\prime})  &  \rightarrow\hat{\Phi}^{+}(r^{\prime
})-\mathrm{i}A^{\ast}\nonumber\\
\hat{\Phi}(r^{\prime})  &  \rightarrow\hat{\Phi}(r^{\prime})+\mathrm{i}A
\label{1.4}%
\end{align}
of\ condensate field operators. This yields
\begin{align}
G_{R}(t)  &  =\left\langle 0\right\vert _{i}\left\langle 0\right\vert \left(
\hat{\Phi}^{+}(r_{b})-\mathrm{i}A^{\ast}\right)  \hat{\Psi}(r_{b}%
)\label{1.5}\\
&  \mathrm{e}^{-\mathrm{i}t\int\mathrm{d}r\mathrm{d}r_{1}\hat{\Psi}%
^{+}(r)\left(  \hat{\Phi}(r)+\mathrm{i}A\right)  V(r-r_{1})\hat{\Psi}%
(r_{1})\left(  \hat{\Phi}^{+}(r_{1})-\mathrm{i}A^{\ast}\right)  }\nonumber\\
&  \left(  \hat{\Phi}(r_{a})+\mathrm{i}A\right)  \hat{\Psi}^{+}(r_{a}%
)\left\vert 0\right\rangle \left\vert 0\right\rangle _{i},\nonumber
\end{align}
where the Green's function arguments $r_{a}$ and $r_{b}$ are omitted.

The mean field approximation implies that one simply ignores the fluctuations
and hence neglects the quantum the field operators $\hat{\Phi}$ and $\hat
{\Phi}^{+}$ acting on the vacuum state as compared to the average macroscopic
amplitudes $A^{\ast}=A=\sqrt{n}$ given by the condensate average density $n$.
The electron tunneling amplitude Eq.(\ref{1.5}) in the momentum representation
thus takes the form%
\begin{align}
G_{R}(t)  &  =n\int\frac{\mathrm{d}^{d}p}{\left(  2\pi\right)  ^{d}%
}\left\langle 0\right\vert _{i}\hat{\Psi}(p)\mathrm{\exp}\left[
\mathrm{i}p(r_{a}-r_{b})\right] \label{1.6}\\
&  \mathrm{\exp}\left[  -\mathrm{i}nt\int\hat{\Psi}^{+}(p)V(p)\hat{\Psi
}(p)\frac{\mathrm{d}^{d}p}{\left(  2\pi\right)  ^{d}}\right]  \hat{\Psi}%
^{+}(p)\left\vert 0\right\rangle .\nonumber
\end{align}

Substituting the explicit expression Eq.(\ref{Interaction}) for $V(p)$ to
Eq.(\ref{1.6}), after straightforward calculations of the matrix element one
obtains%
\begin{equation}
G_{R}(t,R)=\frac{n}{\left(  2\pi\right)  ^{d}}\int\exp\left[  \frac
{-\mathrm{i}t}{p^{2}+1}+\mathrm{i}pR\right]  \mathrm{d}^{d}p, \label{1.7}%
\end{equation}
where $R=r_{a}-r_{b}$ and we have employed the natural units of the problem:
the length unit $1/\kappa\sim0.45\ 10^{-9}m$ and the time unit $\left(
2\pi\right)  ^{d}\hbar/n\Lambda\sim5\ 10^{-4}s$. The numerical values are
given for the case of $3D$ setting and for the condensate density
$n\sim10^{13}cm^{-3}$.

In the natural units, the density $n$ becomes a small dimensionless parameter
$\sim10^{-9}$ that equals to the expectation number of atoms in the volume
$\kappa^{-3}$. The Green's function is thus proportional to $n$. This factor
comes from the initial states and reflects a small probability amplitude to
have the condensate atoms in the vicinities of the point electrodes.
Evidently, this small pre-exponential factor increases considerably when one
integrates over the leads surfaces: for the surface of the order of the
condensate cross-section $S\sim10^{-14}m^{2}$ the enhancement factor can reach
$\sim10^{5}$. One may interpret this factor by saying that the number of
tunneling channels increases proportionally to the lead surface areas. Note
that the transition probability scales as the condensate density square, which
is a natural consequence of the condensate coherence resulting in the
constructive interference of different channels.

\begin{figure}
[ptb]
\begin{center}
\includegraphics[
height=1.3811in,
width=3.3044in
]%
{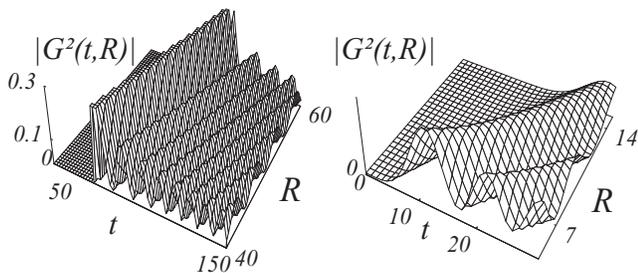}%
\caption{The tunneling probability as a function of scaled time $t$ and scaled
distance between the electrods $R$.\ Results of the saddle point aproximation
on the left. Results of the exact calculation on the right.}%
\label{FIG.2}%
\end{center}
\end{figure}

The large parameter $R$ allows one to evaluate the integral Eq.(\ref{1.7})
with the help of the stationary phase method. The saddle point is given by the
conditions%
\begin{equation}
R=\frac{-2tp_{x}}{\left(  p_{x}^{2}+1\right)  ^{2}};\ p_{y}=p_{z}=0
\label{saddle}%
\end{equation}
Alternatively, this integral can be evaluated numerically, with the allowance
for the fact that the contribution from the asymptotic $p\rightarrow\infty$
corresponding to the Fourier transform of $\delta$ function has to be
subtracted for numerical convergency. The dependence of $\left\vert
G_{R}(t,R)\right\vert ^{2}$ on scaled time and distance is shown in
Fig.\ref{FIG.2} for 1D
The left plot is done with the help of the saddle point approximation, whereas
the right one is an exact calculations based on the representation of the
integral in Eq.(\ref{1.7})
\begin{equation}
G_{R}(t,R)=n\sqrt{\frac{1}{2\pi R}}\sum_{k=0}^{\infty}\frac{\left(
-\mathrm{i}tR/2\right)  ^{k}}{k!(k-1)!}K_{\frac{1}{2}-k}\left(  R\right)
\label{1.9}%
\end{equation}
as a converging series of the Bessel functions $K_{\nu}\left(  R\right)  $,
which results from casting of $\exp\left[  -\mathrm{i}t\left(  p^{2}+1\right)
^{-1}\right]  $ in Taylor series followed by exact integration over $p$ of
each term of the series. For other dimensions just a power of time enters the
denominator of the pre-exponential factor. The saddle point conditions
Eq.(\ref{saddle}) help to understand why the tunneling starts at a time, which
is linearly proportional to $R$: at the real axis $p_{x}$ the combination
$-2p_{x}\left(  p_{x}^{2}+1\right)  ^{-2}$ has the maximum $3\sqrt{3}/8$, and
hence the saddle point locates at the real momentum axis only when the ratio
$R/t$ is less than this number. Note that the contribution of only small $p$
yields the free particle propagator%
\begin{equation}
G_{R}(t,R)=\frac{n}{\left(  -4\mathrm{i}\pi t\right)  ^{d/2}}\exp\left[
-\mathrm{i}t-\mathrm{i}\frac{R^{2}}{4t}\right]  . \label{frreprop}%
\end{equation}

\begin{figure}
[ptb]
\begin{center}
\includegraphics[
height=1.7824in,
width=2.7337in
]%
{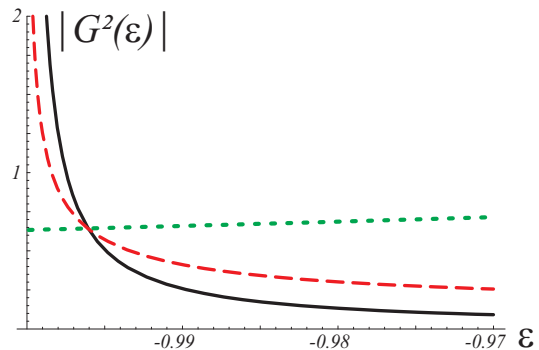}%
\caption{Scaled transition probability $\left\vert R^{d-1}G_{R}(\varepsilon
,R)/n\right\vert ^{2}$ for the mean field model as a function of the energy
$\varepsilon$.\ $1D$-solid line, $2D$ -dash line, $3D$ -dot line. }%
\label{FIG.1}%
\end{center}
\end{figure}

As it has already been mentioned in the beginning of this Section, the time
dependent approach is not directly relevant to case of large size of the
tunneling wave packets, whereas the energy representation only has a physical
meaning. By applying the Fourier transformation in time $\int_{0}^{\infty
}\mathrm{d}t\ldots\exp[-\mathrm{i}\varepsilon t]$ to Eq.(\ref{1.7}) and
neglecting the singular terms at $R=0$ we find the energy dependent transition
probability amplitudes
\begin{align}
G_{R}(\varepsilon,R)|_{1D}  &  =\frac{\mathrm{i}n}{2\varepsilon^{2}}%
\sqrt{\frac{\varepsilon}{\varepsilon+1}}\exp\left(  -R\sqrt{\frac
{\varepsilon+1}{\varepsilon}}\right) \label{1.10}\\
G_{R}(\varepsilon,R)|_{2D}  &  =\frac{\mathrm{i}n}{2\pi\varepsilon^{2}}%
K_{0}\left(  R\sqrt{\frac{\varepsilon+1}{\varepsilon}}\right) \nonumber\\
G_{R}(\varepsilon,R)|_{3D}  &  =\frac{\mathrm{i}n}{4\pi R\varepsilon^{2}}%
\exp\left(  -R\sqrt{\frac{\varepsilon+1}{\varepsilon}}\right)  ,\nonumber
\end{align}
which yield probabilities depicted in Fig.\ref{FIG.1}.
Note that in the dimensional units, the point $\varepsilon=0$ corresponds to
the electron affinity energy, and the point $\varepsilon=-1$ is shifted with
respect to this value by $-n\Lambda/\left(  2\pi\right)  ^{d}$. The physically
meaningful domain corresponds to $\varepsilon$ close to $-1$ where the typical
momentum is small as compared to $\kappa$, which is consistent with the free
particle expression Eq.(\ref{frreprop}). Moreover, one finds a smaller cutoff
momentum $p_{c}\sim n^{-d}$ for the electron by imposing the requirement that
the phase space integral $\frac{\mathcal{V}}{\left(  2\pi\right)  ^{d}}%
\int^{p_{c}}\mathrm{d}^{d}p$ equals the total number of the electron quantum
states, that is the number of atoms $\mathcal{N}$. Going beyond this momentum
is inconsistent with the formfactor approximation Eq.(\ref{Formfactor}). Note
that the correlation effects in the interatomic distance discussed in
Ref.\cite{Mazets}

\section{The role of quantum density fluctuations\label{IV}}

The mean field picture gets drastically changed when the quantum fluctuations
of the condensate density are taken into account. The fluctuations cause
scattering of the electron in the course of tunneling, while the energy and
the momentum transfer associated with this scattering induce quantum
transitions in the condensate. Analytical consideration of this regime is
technically rather involved and rely on the functional integration over the
regular and Grassmann fields. This method, known as $\sigma$-model, is widely
employed for description of disordered systems, and we just adapt it to for
case of averaging over the condensate quantum state instead of the ensemble
averaging, for which it has been originally developed.

We start with the technically more simple time-dependent version of
Eq.(\ref{ProbTun}) for the tunneling probability%

\begin{align*}
&  \int\psi_{fn}(\left\{  r_{n}\right\}  )G_{A}(\left\{  r_{n}\right\}
,\tau,r_{a},r_{b})\psi_{in}^{\ast}(\left\{  r_{n}\right\}  )d^{\mathcal{N}}r\\
&  \times\int\psi_{fn}^{\ast}(\left\{  r_{n}\right\}  )G_{R}(\left\{
r_{n}\right\}  ,t,r_{b},r_{a})\psi_{in}(\left\{  r_{n}\right\}
)d^{\mathcal{N}}r,
\end{align*}
and construct a generating function%
\begin{equation}
P(t,\tau,z)=\sum_{fn}\left\langle in\right\vert G_{A}(\tau)\left\vert
fn\right\rangle e^{z\delta\mathcal{N}}\left\langle fn\right\vert
G_{R}(t)\left\vert in\right\rangle , \label{GF}%
\end{equation}
where $\delta\mathcal{N}$ denotes the difference in the occupation numbers
between the initial $\left\vert in\right\rangle $and the final $\left\vert
fn\right\rangle $ states of the condensate. This function allows one to find
several important physical quantities. In particular, the Fourier transform of
the generating function
\[
P(\eta,z)=\int\limits_{-\infty}^{\infty}P(t,\tau,z)\exp[-\mathrm{i}\eta\left(
t-\tau\right)  ]\mathrm{d}t\mathrm{d}\tau
\]
at $z=0$ equals to the net electron tunneling probability at the energy $\eta$
with the allowance of all possible changes of the condensate quantum state
induced by the tunneling. The derivative $\partial_{z}P(\eta,z)|_{z\rightarrow
0}$ gives the number of atoms that have left the condensate as a result of the
tunneling. The probability of tunneling with no condensate transition
corresponds to $P(\eta,-\infty)$.

The explicit form of the generating function Eq.(\ref{GF}) corresponding to
the initial state of condensate Eq.(\ref{Initial}) reads
\begin{align}
P(t,\tau,z)  &  =\left\langle 0\right\vert \left\langle 00\right\vert
_{i}\widehat{U}_{d}^{-1}\overline{\hat{\Psi}}(r_{a})\hat{\Phi}^{+}%
(r_{a})\mathrm{e}^{\mathrm{i}\tau\widehat{H}_{A}}\label{2.1}\\
&  \hat{\Phi}(r_{b})\overline{\hat{\Psi}}^{+}(r_{b})\widehat{U}_{d}%
\mathrm{e}^{z\widehat{\mathcal{N}}}\widehat{U}_{d}^{-1}\hat{\Phi}^{+}%
(r_{b})\hat{\Psi}(r_{b})\nonumber\\
&  \mathrm{e}^{-\mathrm{i}t\widehat{H}_{R}}\hat{\Phi}(r_{a})\hat{\Psi}%
^{+}(r_{a})\widehat{U}_{d}\left\vert 00\right\rangle _{i}\left\vert
0\right\rangle ,\nonumber
\end{align}
where\ $\widehat{U}_{d}$ is the displacement operator Eq.(\ref{displacement}),
$\widehat{\mathcal{N}}=\int\mathrm{d}r\hat{\Phi}^{+}(r)\hat{\Phi}(r)$ is the
number of particles operator,
\begin{align}
\widehat{H}_{A}  &  =\int\mathrm{d}r\mathrm{d}r_{1}\overline{\hat{\Psi}}%
^{+}(r_{1})\hat{\Phi}(r_{1})V(r-r_{1})\hat{\Phi}^{+}(r)\overline{\hat{\Psi}%
}(r)\nonumber\\
\widehat{H}_{R}  &  =\int\mathrm{d}r\mathrm{d}r_{1}\hat{\Psi}^{+}(r_{1}%
)\hat{\Phi}(r_{1})V(r-r_{1})\hat{\Phi}^{+}(r)\hat{\Psi}(r) \label{HAHR}%
\end{align}
are Hamiltonians for the advanced and retarded Green's functions respectively,
field operators with bars correspond to the advanced Green's function, and the
vacuum state $\left\vert 00\right\rangle _{i}$ is the direct product of the
vacuum states for $\hat{\Psi}$ and $\overline{\hat{\Psi}}$.

In the first line of Eq.(\ref{2.1}) one recognizes the combination
$\left\langle in\right\vert G_{A}(\tau)$, \ in the last line of this equation
stands the combination $G_{R}(t)\left\vert in\right\rangle $, while the middle
line gives the expression for $\sum_{fn}\left\vert fn\right\rangle
\mathrm{e}^{z\delta\mathcal{N}}\left\langle fn\right\vert .$ Let us explain
the physical meaning of the last combination in more detail. The initial
coherent state of condensate is constructed by applying the displacement
operator Eq.(\ref{displacement}) to the condensate vacuum (no atoms only
vacuum fluctuations). If during the electron tunneling no change of this state
occur, the inverse displacement operator restores the vacuum and the operator
$\exp z\int\hat{\Phi}^{+}(r)\hat{\Phi}(r)\mathrm{d}r$ makes no action. If a
change occur, the inverse displacement does not restore the vacuum, but yields
a state $\left\vert x\right\rangle $ with the expectation value $\left\langle
x\right\vert \int\hat{\Phi}^{+}(r)\hat{\Phi}(r)\mathrm{d}r\left\vert
x\right\rangle $ equal to average number $\delta\mathcal{N}$ of the atoms
extracted from the condensate. Hence, application of the operator $\exp
z\int\hat{\Phi}^{+}(r)\hat{\Phi}(r)\mathrm{d}r$ yields the required factor
$\mathrm{e}^{z\delta\mathcal{N}}$.

For description of the effect of atomic density fluctuations on the electron
tunneling probability, one needs to separate the condensate and the electron
motions. There is a standard way to do it with the help of functional
integration. One represents the evolution operator for a bipartite system as a
product of two commuting operators each of which acts on only one of the
parts. Each of these operators depends however on a common function, a sort of
external field, which serves as a variable for the subsequent functional
integration. In this way one can first calculate the required matrix elements
for the initial coherent condensate state of the atomic part and find as the
result a weight functional for the electron motion.\ Thus obtained weight
functional is an analog of the distribution functional $W(U)$ in
Eq.(\ref{Tuu}) of random potentials employed in the theory of disordered
conductors, and therefore further calculations can be done by the analogy, --
with the help of a technique known as the non-linear $\sigma$-model.

\begin{figure}
[h]
\begin{center}
\includegraphics[
height=1.7028in,
width=2.8141in
]%
{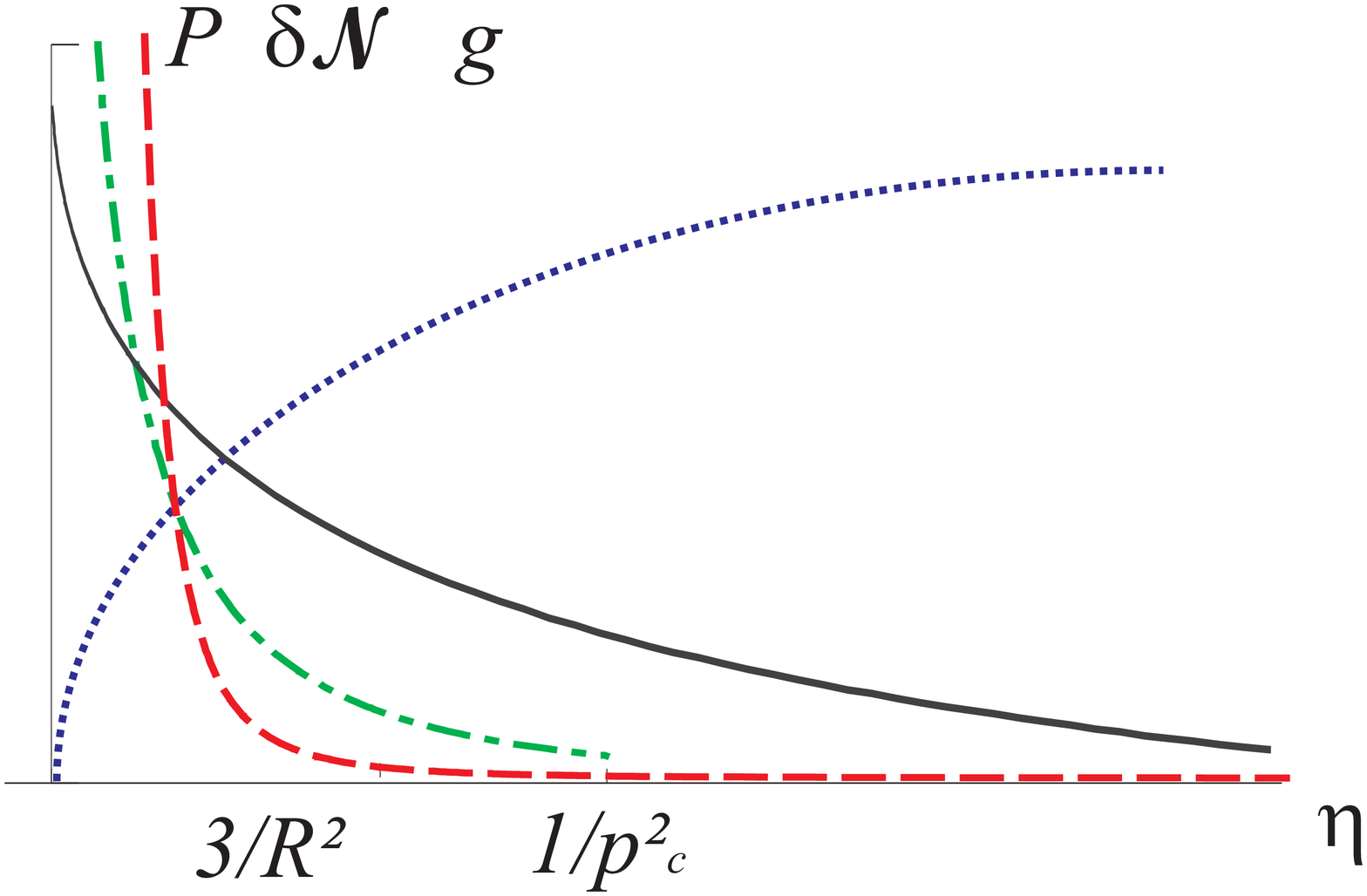}%
\caption{ Effect of the condensate density fluctuations on the tunneling
probability $P$. The probability (solid line) changes with the distance $R$
and the energy $\eta$ according the law $\exp\left(  -2R\sqrt{\eta}\right)  $
and differs from the result of the mean field model (dash-dot line). The
quantum state \ density $g$ and the number $\delta\mathcal{N}\propto$
$\eta^{-(3+d)/2}$ of the atoms kicked out from the condensate for $2D$ case
are shown by dot line and dash line, respectively. The dependences are shown
not to scale, and the energy $\eta=0$ here corresponds to the band edge}%
\label{FSM}%
\end{center}
\end{figure}

Details of these rather \ long analytic calculations are given in the Appendix
Sect.\ref{1}. The final expression for the generating function has an explicit
form%
\begin{align}
&  P(\zeta,\eta,z)=-2G_{A}\left(  R\sqrt{\frac{\eta+\mathrm{i}q_{s}}{2}%
}\right)  G_{R}\left(  R\sqrt{\frac{\eta-\mathrm{i}q_{s}}{2}}\right)
\nonumber\\
&  ~\operatorname{Re}\int_{1}^{\infty}\mathrm{d}Z\int_{-1}^{1}~\mathrm{d}%
w\mathrm{e}^{-\mathcal{N}2\left(  1-\mathrm{e}^{-z}\right)  q_{s}^{2}\left(
Z^{2}-w^{2}\right)  -\mathrm{i}2\mathcal{N}\zeta q_{s}\left(  Z-w\right)  },
\label{2.02}%
\end{align}
in terms of the electron energy $\eta=\frac{\varepsilon+\xi}{2}$ and the
frequency $\zeta=\frac{\varepsilon-\xi}{2}$. Here $q_{s}=\operatorname{Im}%
\sqrt{\eta^{2}-2}$ is a parameter related to the electron states density
profile, and Green's functions Eq.(\ref{2.02}) given explicitly in
Eq.(\ref{GRANF}) in fact coincide with that given by Eq.(\ref{1.10}) where the
arguments $\sqrt{\frac{\varepsilon+1}{\varepsilon}}$are simply replaced by the
combination $\mathrm{i}\left(  \eta\pm\mathrm{i}q_{s}\right)  ^{1/2}/\sqrt{2}%
$. Results for the net tunneling probability $\mathcal{P}(t\rightarrow
\infty,\eta)=P(0,\eta,0)$ and the number $\delta\mathcal{N}(t\rightarrow
\infty,\eta)=P_{z}^{\prime}(0,\eta,0)$ of atoms extracted from the condensate
are depicted in Fig.\ref{FSM} for $2D$ case along with the probability
suggested by the mean field model.
As it will now be explained, the dimensionality of the problem plays here just
a marginal role, and does not affect the conductance property in the domain,
which gives the main contribution to the tunneling probability.

Let us give a qualitative picture of the tunneling process emerging from the
analytical consideration. The mean field model considers atoms as uniformly
distributed in space and relies on the atomic density smeared over the space.
The tunneling electron mobility then results from the dispersion law
Eq.(\ref{Interaction}), which close to the lower spectrum edge can be
considered as the kinetic energy of a free particle with a large effective
mass, whereas the potential energy is ignored. However within the
Born-Oppenheimer approximation, the electron potential energy created by
frozen atoms, on the contrary, is a strongly irregular function composed by
$\delta$-like spikes placed at generic atomic positions in space. Were these
atoms be indeed fixed in space, the tunneling electron would experience the
Anderson localization effect resulting from the quantum interference of many
different multiple scattering trajectories. At the low-energy edge, the
electron state density spectrum would correspond to short Anderson
localization lengths, and the tunneling would be strongly suppressed. The
difference between this case and the case under consideration is in the fact
that the frozen atoms are not fixed, and even without changing their average
space positions they still can change quantum states by changing their momenta
as the result of the multiple electron scattering. This destroys the electron
trajectory interference and the related influence of the dimensionality of the
problem. It is therefore not surprising that the result can be found in the
framework of zero-dimensional $\sigma$-model, while the role of the
dimensionality is reduced to purely geometric factors. We note that in such a
situation the Green's functions keep almost the same form of the coordinate
dependence just experiencing a shift of the energy scale and acquiring
imaginary parts of the energy arguments corresponding to the rate of quantum
transitions among the states of condensate.

Suppression of the Anderson localization for an electron tunneling through the
condensate can also be viewed by analogy to the quantum molecular transitions.
In a molecule with fixed atomic positions, the electronic energies take
discrete values thus forming electronic terms, which can be seen as a
consequence of the interference of electron trajectories. A weak external
fields is unable to induce purely electronic transitions without involving the
nuclear motion, unless the resonance conditions are satisfied, - the
interference prevents electrons from changing of the quantum states. The
resonances implies exact tuning of the\ field frequency, which for given
positions of atoms can be achieved only for a specific choice a pair of the
electronic states. However for the polyatomic molecules, when one takes into
account the nuclear motion with a very dense vibrational spectrum, the
resonance conditions can be matched for many electronic terms at once. In
other words, the quantum interference of the purely electronic motion gets
destroyed by the vibrations. The transition probability is then governed by
the Frank-Condon factors resulting from the overlaps of the vibrational eigen
functions corresponding to different electronic terms. For the condensate, the
role of the Frank-Condon factors play the matrix elements Eq.(\ref{ProbTun}),
whereas the role of the vibrational energy quantum numbers play the second
quantization occupation numbers of the condensate modes.

\begin{figure}
[h]
\begin{center}
\includegraphics[
height=2.143in,
width=3.2569in
]%
{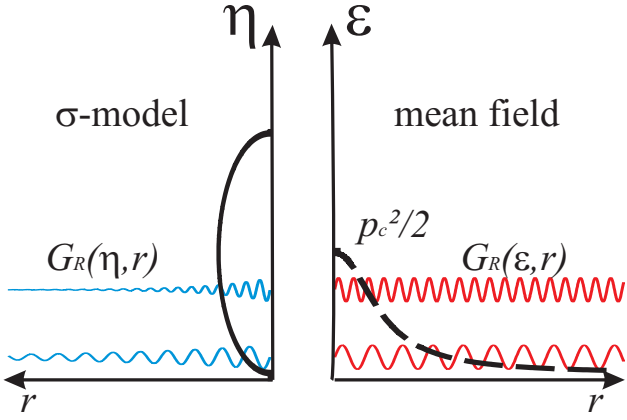}%
\caption{State densities and Green's functions for the mean field $1D$ model
and with allowance of quantum flucuations in $0D$ $\sigma$-model. Solid line
represents the state density with disorder, while the dashed line corresponds
to the mean field model. These profiles can be refered to as the electron
conductance bands of the condensate.}%
\label{FTT1}%
\end{center}
\end{figure}

Let us now compare and contrast the results of the mean field model and the
$0D$ $\sigma$-model, illustrated in Fig.\ref{FTT1}. Near the bottom spectrum
edge the density of states in the mean field approximation correspond to that
of free particle, while the allowance of the quantum fluctuations results in a
Wigner semicircular density distribution.
In both cases the tunneling probability is proportional to the product of the
coordinate dependent advance $G_{A}$ and and retarded $G_{R}$ Green's
functions. However, for the mean field model, they both are harmonic functions
of the distance $r$ and correspond to the free particle motion, while the
allowance for density fluctuations makes $G_{A}$ and $G_{R}$ exponentially
decaying with $r$. In the mean field model, $G_{A}\left(  \xi,r\right)  $ and
$G_{R}\left(  \varepsilon,r\right)  $ depending on different energy variables
$\xi$ and $\varepsilon$ , respectively, after the inverse Fourier
transformation yield two wave packet propagators related by the complex
conjugation condition $G_{A}\left(  t,r\right)  =G_{R}^{\ast}\left(
t,r\right)  $. The result of $\sigma$-model is different: $G_{A}\left(
\eta,r\right)  $ and $G_{R}\left(  \eta,r\right)  $ being exponentially
decreasing complex conjugate functions of $r$ both depend on the same energy
variable $\eta$, while the frequency dependence of the type $-\frac{\sin
^{2}\left(  g(\eta)\zeta\right)  }{\left(  g(\eta)\zeta\right)  ^{2}}$ factors
out and yields a linear time dependence as a common cofactor of the product
$G_{A}G_{R}$. Here $g(\eta)$ is the density of states at the energy $\eta$
(see Eq.(\ref{LinTProd})). The situation thus resembles behavior of the
tunneling evanescent waves under a potential barrier in the regular
formulation of this quantum-mechanical problem, although here the transition
probability
\begin{figure}
[h]
\begin{center}
\includegraphics[
height=2.0003in,
width=3.3295in
]%
{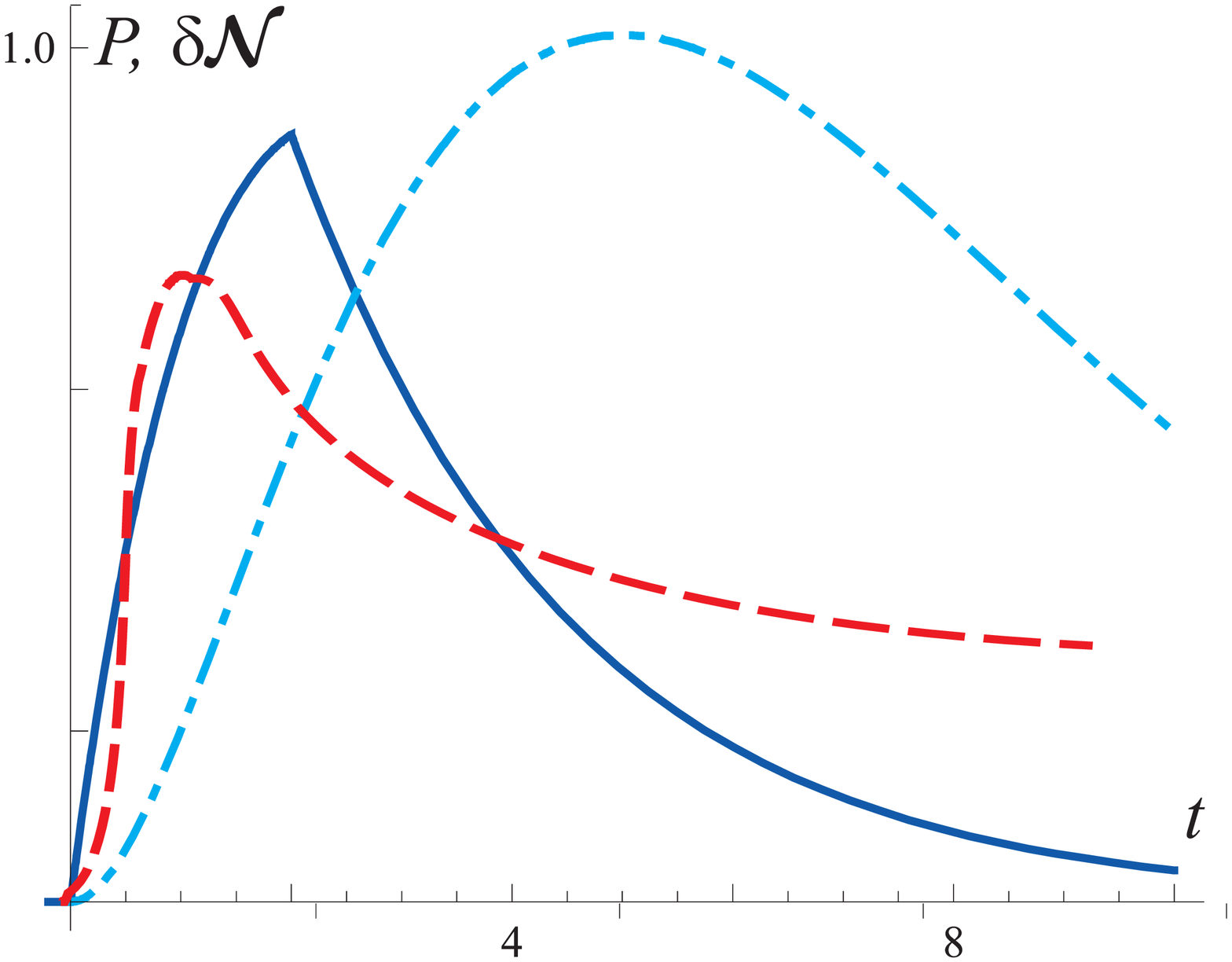}%
\caption{Time dependence of the tunneling probability $P$ with the allowance
of the fluctuations (solid line) and for the mean field model (dash line). The
rate of the condensate atoms loss $\delta\mathcal{N}_{t}^{\prime}$ (dash-dot
line). Curves not to scale.}%
\label{TTN}%
\end{center}
\end{figure}
linearly increases with time till it reaches the saturation value, as shown in
Fig. \ref{FourierCoff} in the Appendixs. Allowance for the finite life time of
the eletron in the condensate associated with it's returns to the leads,
result in a certain transformation of this dependence shown in Fig.\ref{TTN},
although it still differs essentially from the tunneling decay law, which is
usually an exponential one.

\begin{figure}
[h]
\begin{center}
\includegraphics[
height=1.9078in,
width=3.026in
]%
{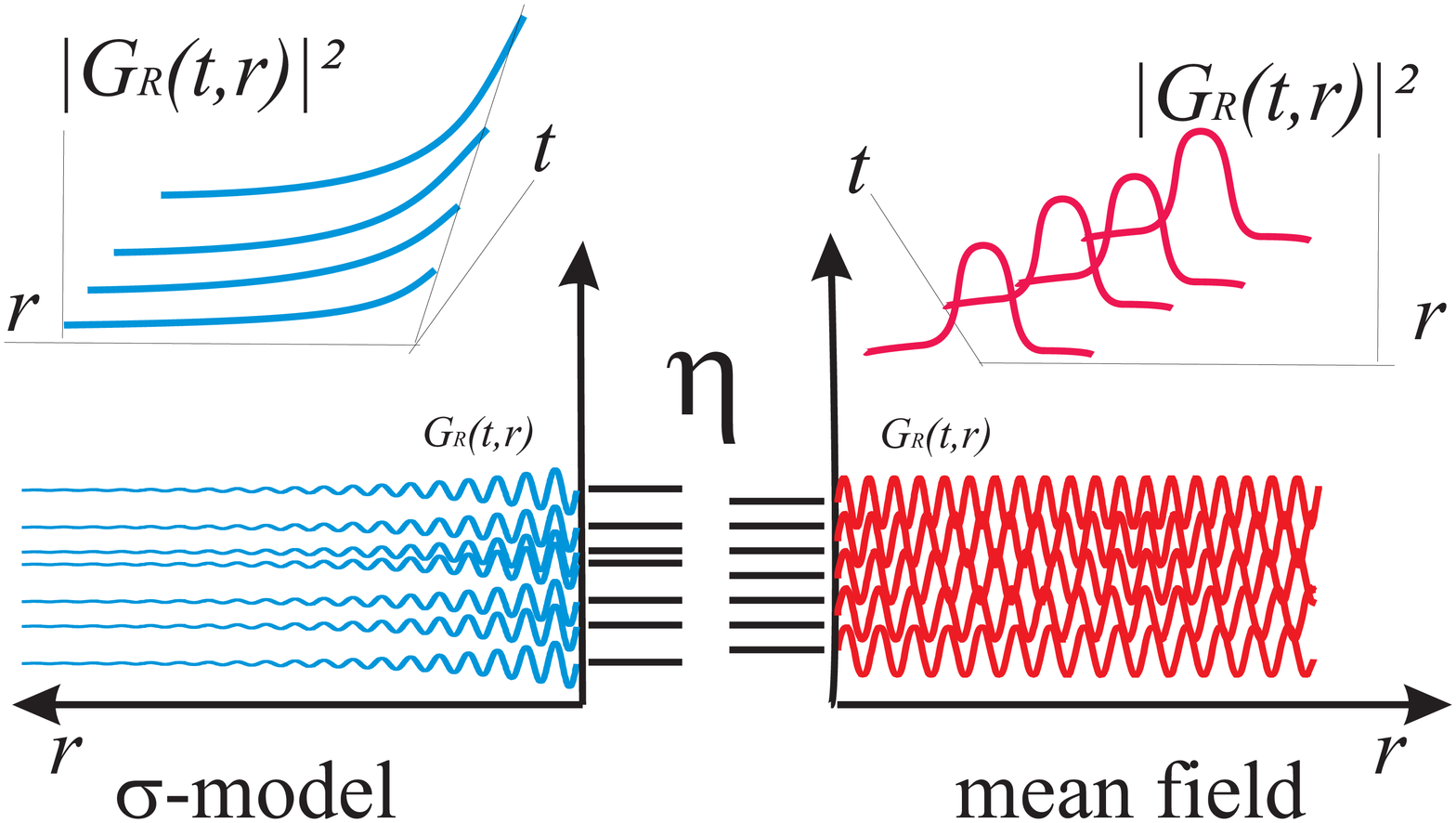}%
\caption{Regular and almost equidistant spectrum of the electron in the mean
field model (to the right) corresponding to the plane wave eigenfunction
tesults in propageting wave pachets. Irregular spectrum (to the left) with GUE
statistics and localized eigen functions results in the evanecent wave that
linearly increases during a finite interval of time, till the Heisenberg time,
when it saturates.}%
\label{ftt2}%
\end{center}
\end{figure}

The linear time dependence is typical of random ensembles of unitary matrices,
and it reveals statistical nature of quantum fluctuations hidden in the
coherent states of the atomic condensate. This is illustrated in
Fig.\ref{ftt2}.
Note that total tunneling probability given by the integral of the linear time
dependence corresponds to the limit $\zeta\rightarrow0$ in the Fourier
transform $\frac{\sin^{2}\left(  g(\eta)\zeta\right)  }{\left(  g(\eta
)\zeta\right)  ^{2}}$ Eq.(\ref{LinTProd}) and always gives unity. However, the
time at which the tunneling reaches the stationary value equals the Heisenberg
time given by the state density $g(\eta)$, and thus the smaller is the state
density the shorter is the saturation time. This results is the consequence of
the random matrix theory where all spectral properties and their time
dependent counterparts are governed exclusively by the local density of states.

One has to emphasize an important qualitative difference between the electron
moving in a disordered potential and the electron tunneling through the
condensate. For the condensate, one of the channels contributing the electron
transport should be avoided, if possible. This is the channel when the
electron makes not a virtual but a real transition to some point in between
the leads, which implies the electron charge localization in this point and
hence provokes complete destruction of the condensate. It does not contribute
to the process of tunneling. In contrast, this processes does give the
dominating contribution to electronic diffusion in disordered potentials, and,
evidently, is influenced by the space dimensionality. In other words, the
tunneling regime requires the condensate size being of the order of the
electron mean free path in the corresponding diffusion setting. Due to this
reason, utilization of $O$ dimensional $\sigma$-model for the description of
the tunneling looks as a reasonable approach.

\section{The role of atomic and the lead material properties, and the
tunneling efficiency in numbers\label{V}}

In order to discuss the possibility and requirements for an experimental
observation of the electron tunnelling through a condensate, we now present
the typical probability and time scales of the tunneling process in the
dimensional units along with the related spectral characteristics. Since the
objective is to describe the electron transition from one lead to the other,
the corresponding probability amplitudes should be given in terms of
electronic states, and hence the ionic point-to-point Green's functions
Eqs.(\ref{1.7} ,\ref{GRANF}) have to be projected to the lead electron wave
functions $\varphi_{l}^{\ast}(r_{b})$ and $\varphi_{l}(r_{a})$ with the help
of Eq.(\ref{Formfactor}), which relates position $r$ of the negative ion with
the wave function $\varphi(r-r^{\prime})$ of the attached electron. The main
contribution to the projection comes from the evanescent tails of $\varphi
_{l}$ integrated over two volumes $\mathrm{d}z_{a}\mathrm{d}S_{a}$ and
$\mathrm{d}z_{b}\mathrm{d}S_{b}$ near surface of the lead \ $a$ of area
$S_{a}$ and near surface of the lead $b$ of area $S_{b}$, respectively.

Among all possible $\varphi_{l}$, the maximum overlap in the directions
$z_{a;b}\perp S_{a;b}$ eventually comes from the electronic surface states
$\varphi_{sl}(r_{a;b})$, while the overlap integrals%
\begin{align*}
\phi(r_{a})  &  =\int\mathrm{d}r_{a}^{\prime}\varphi^{\ast}(r_{a}%
-r_{a}^{\prime})\varphi_{sl}(r_{a}^{\prime})\\
\phi^{\ast}(r_{b})  &  =\int\mathrm{d}r_{b}^{\prime}\varphi(r_{b}%
-r_{b}^{\prime})\varphi_{sl}^{\ast}(r_{b}^{\prime})
\end{align*}
playing role of the initial and the final ionic wave functions are
well-localized in the vicinity of the surfaces. If we assume a typical size of
the evanescent tails of the order of $\kappa^{-1}$ and take the lead areas big
enough $S_{a}\sim S_{b}\gg\kappa n^{-1}$, which implies many condensate atoms
on average located at a distance $<\kappa^{-1}$ from the surface, the
pre-exponential factor $n$ in Eqs.(\ref{1.7} ,\ref{GRANF}) can be replaced by
a large parameter $N_{c}=Sn\kappa^{-1}$ that has a physical meaning of the
number of tunneling channels. Note that "on average" is understood here not in
the statistical but in the quantum sense. Different tunneling channels
therefore interfere, whereas the results of such an interference for $3D$ are
given by the integrals%

\begin{align*}
G_{R}  &  =\int\mathrm{d}S_{a}\mathrm{d}S_{b}\frac{\mathrm{i}N_{c}\phi^{\ast
}(r_{b})\phi(r_{a})}{4\pi\kappa\left\vert r_{a}-r_{b}\right\vert }%
\mathrm{e}^{-\mathrm{i}\left(  r_{a}-r_{b}\right)  \kappa\sqrt{\varepsilon}%
},\\
G_{R}  &  =\int\mathrm{d}S_{a}\mathrm{d}S_{b}\frac{N_{c}\phi^{\ast}(r_{b}%
)\phi(r_{a})}{4\pi\kappa\left\vert r_{a}-r_{b}\right\vert }~\mathrm{e}%
^{-\mathrm{i}\left(  r_{a}-r_{b}\right)  \kappa\left(  \frac{1}{\sqrt{2}%
}-\mathrm{i}\frac{\sqrt{\eta}}{\sqrt[4]{2}}\right)  },
\end{align*}
corresponding to the mean field and the $\sigma$ models, respectively.\ The
amplitudes $\phi$ are now defined at the surfaces and are normalized by the
condition $\int\phi^{\ast}\phi\mathrm{d}S=1$, while the energies $\varepsilon$
and $\eta$ are chosen close to the corresponding condensate conducting band
edges. One can move the big distance $R=\left\vert r_{a}-r_{b}\right\vert $ in
front of the integrals and obtain%
\begin{align}
G_{R}(\varepsilon,R)  &  =\mathrm{i}N_{c}\frac{\phi^{\ast}(\kappa
_{\shortparallel R}\sqrt{\varepsilon})\phi(\kappa_{\shortparallel R}%
\sqrt{\varepsilon})}{4\pi\kappa R},\label{TUP}\\
G_{R}(\eta,R)  &  =N_{c}\frac{\phi^{\ast}(\frac{\kappa_{\shortparallel R}%
}{\sqrt{2}})\phi(\frac{\kappa_{\shortparallel R}}{\sqrt{2}})}{4\pi\kappa
R}~\mathrm{e}^{-\frac{\sqrt{\eta}}{\sqrt[4]{2}}R\kappa},\nonumber
\end{align}
where the surface functions are now in the momentum representation.

One sees that the tunneling probability depends on the surface electron wave
functions that, in turn, are governed by the band structure of the bulk
material. For a broad-band semiconductor with a low electric polarizability,
the surface electron may form a Rydberg-like band spectra bounded to the
surface with an energy $E_{s}$ due to the presence of the weak image
force.\ For a small $E_{s}$, the wave functions corresponding to these surface
bands are rather extended in the direction normal to the surface. In this
case, the number of the channels involved in tunneling yet increases. However,
it may require compensation for eventual energy mismatches between the band
energies and the atomic affinity, which in principle can be done by a proper
periodic external perturbation. One can also argue that the materials with low
electric susceptibility have less chances to destroy the condensate by the Van
der Waals interaction.

Detailed consideration of the optimum choice of the material is a separate
task deserving a special consideration. Here we will not dwell on this
interesting issue and simply take the factors $\phi$ in Eq.(\ref{TUP}) of the
order of unity. Still it worth mentioning that different models address
different types of the surface wave functions. The mean field model by
considering the tunneling electron as a free propagating particle of a large
mass imposes the phase matching conditions between the plane waves of the
tunneling and the surface electrons. Due to the difference in effective masses
of these electrons, the conditions are fulfilled near the bottoms of the
relevant bands, -- the tunneling band and the surface band corresponding to
small momenta. The bottoms of the bands thus have to coincide in this case.
Since now the energy only enters the arguments of $\phi$, the tunneling
dynamics is governed by the shape of the surface state density profile near
zero momenta. In contrast, the $\sigma$- model for the fluctuations results in
irregular localized eigenfunctions of the tunneling band that are correlated
only at the atomic scale of sizes, and therefore the main contribution comes
from the surface states with the momenta of the order of $\kappa$. The time
dependence is formed by the tunneling band level statistics. Note that in
order to be consistent within the tunneling regime approximation, in both
cases we have to consider the tunneling times as short compared to the
duration of the surface electron wave packet.

Let us now estimate the numerical values for the tunneling probability and the
typical tunneling time at the example of $Cs$. Fot typical dimensionless
energies $\eta\sim1/R$, both models yield%
\begin{equation}
P\simeq\left(  \frac{N_{c}}{4\pi\kappa R}\right)  ^{2}\simeq\left(  \frac
{Sn}{4\pi\kappa^{2}R}\right)  ^{2} \label{pR}%
\end{equation}
that for $\kappa=2.26\ 10^{9}\left[  m^{-1}\right]  ,$ $n\sim10^{13}cm^{-3}$,
$S\sim10^{-14}m^{2}$, $R\sim10^{-7}m$ amounts to
\[
P\simeq2.4~10^{-16}.
\]
Note that the tunneling probability Eq.(\ref{pR}) is not associated with the
inverse time of the tunneling, -- it is just a geometrical factor which gives
the branching ratio of the small probability to make the transition to another
lead and a much bigger probability to return back to the same lead. The
tunneling time corresponding to this density is definitely superior to the
time of tunneling between two atoms at a distance $n^{-1/3}$, which for
$n\sim10^{13}cm^{-3}$ amounts to a completely unrealistic number%
\begin{equation}
T_{iat}=\frac{E_{a}}{\hbar}\mathrm{e}^{-\kappa n^{-1/3}}\simeq10^{441}\sec.
\label{TUUT}%
\end{equation}
Therefore at the condensate densities $n\sim10^{13}cm^{-3}$ experimentally
accessible at the moment, the tunneling electro-conductance is absolutely impossible.

However, $T_{iat}$ exponentially decreases with the interatomic distance and
gets in the reasonable domain only for the densities $n\sim10^{18}cm^{-3}$
corresponding to the interatomic distance of $100\mathring{A}$, where one
finds the time%
\[
T_{iat}\simeq7~10^{-5}\sec
\]
and the correspondent tunneling probability%
\[
P\simeq2.4~10^{-6}.
\]
Therefore in the regime of dense condensate with interatomic distance of the
order of the scattering length $a\sim50\mathring{A}$, the tunneling approaches
a realistic limit. In the mean-field approximation, the typical time for
tunneling through all the condensate volume is of the order of $\kappa
RT_{iat}\simeq16\mathrm{msec}$. In order to get an idea about the
corresponding tunneling current, one has to multiply the probability by the
electron charge, and divide by the typical tunneling time, which again yields
a small number $J=2.3~10^{-23}A$ that can be increased up to $4~10^{-18}A$ by
increasing the atomic density up to the limit $n\sim~10^{19}cm^{-3}$ suggested
by the scattering length. Evidently, this regime remains beyond the limits of
experimental feasibility and still implies an individual electron counting.
Unfortunaly such densities are unrealistic due to three-body recombination and
dimers formation the condensate gets destroyed at a much shorter time scale
and at much smaller densities..

Still the situation is not completely hopeless, since the electron affinity
enters the exponent in Eq.(\ref{TUUT}). Therefore, the tunneling probability
and the corresponding current increase drastically for the atoms with low
electronic affinity, such as $Ca$, where $E_{a}\simeq20meV$\cite{Ca}%
\cite{Ca1}. At the same atomic density, the current is of the order of
$J\simeq6~10^{-2}pA$, the tunneling time $\kappa RT_{iat}\simeq6~10^{-12}%
\mathrm{sec}$, and the tunneling probability $P\simeq10^{-1}$, which can be
considered as the most optimistic estimation of the tunneling process
efficiency. Even at the condensate density $n\sim~10^{15}cm^{-3}$, one finds
$P\simeq10^{-9}$, while the tunneling time remains at the observation limit
$\kappa RT_{iat}\simeq10^{-2}\mathrm{sec}$. Moreover, estimations show that at
this densities the three-body recombination rate $K_{3}n^{2}$ with
$K_{3}\symbol{126}0.4\ 10^{-31}\mathrm{cm}^{6}/\mathrm{sec}$ is of the order
of $0.04\ \mathrm{sec}^{-1}$ , which is still small enough to be ignored.
However, condensation of rare-earth gases is a much more challenging task,
although the possibility to control the scattering length in a wide region
$a\sim10^{2}-10^{4}\mathring{A}$ by making use of Feichbach resonances has
been studied theoretically \cite{Range} and addressed experimentally
\cite{Range1}.

\begin{figure}
[h]
\begin{center}
\includegraphics[
height=1.8697in,
width=3in
]%
{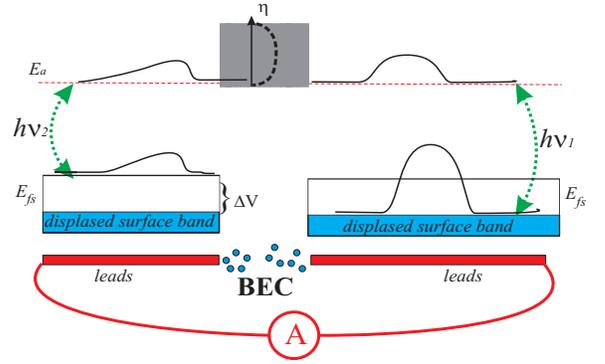}%
\caption{Photon-assisted tunneling of an electron from the surface band to the
states of the negative ions followed by the tunneling through the condensate
and the photoin assistant transition to the surfase state of the other
lead.\ The frequency detuning $\nu_{1}-\nu_{2}$ between the fields acting on
the right and on the left leads emmulate the voltage difference $\Delta V$.
The Fermi energy $E_{sf}$ corresponds to the surface states.}%
\label{MaserTun}%
\end{center}
\end{figure}

Let us now turn to the spectral properties of the tunneling process, where the
difference between the mean field and the $\sigma$-model is the most
pronounced. From the typical size of the tunneling time one sees that the
spectral characteristics manifest themselves in the radio frequency domain, in
other words they correspond to the energy scale of $10^{-9}-10^{-12}eV$, which
practically excludes the experimental studies based on the naive electric
circuit setting shown in Fig.\ref{CoCoWaveFunction}. However, it becomes
possible to employ high-Q sources of electromagnetic radiation at the
frequency domain required for the compensation of the difference between the
surface band energy and the electron affinity of atoms and emulate in this way
small voltage differences applied to the leads, as it is illustrated in
Fig.\ref{MaserTun}.%
This setting allows one to trace the spectral dependence of the tunneling
probability and thus to distinguish between the mean field and the $\sigma
$-model predictions.

\begin{figure}
[h]
\begin{center}
\includegraphics[
height=2.2451in,
width=3.2526in
]%
{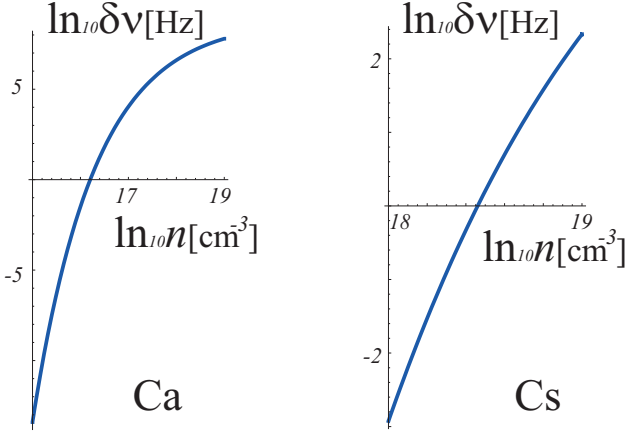}%
\caption{The inhomogeneous linewidth $\delta\nu$ as a function of the
conensate density.}%
\label{Linewidth}%
\end{center}
\end{figure}

According to Eqs.(\ref{TUP}), the tunneling probability for the mean field
model is energy independent as long as we remain near the lowest edge of the
tunneling band and discard the effect of the surface band structure. Allowance
for the density fluctuations results in an exponential square root energy
dependence%
\begin{equation}
P(\eta,R)=\frac{1}{\mathcal{N}\sqrt{2\eta}}\left(  \frac{N_{c}}{4\pi\kappa
R}\right)  ^{2}~\mathrm{e}^{-2\sqrt{2\eta}R\kappa}\nonumber
\end{equation}
which does not depend of the frequency $\zeta$ , thus it is not associated
with the inverse of a tunneling time and can therefore be interpreted as an
inhomogeneous spectral line. The line width $\delta\eta=1/\left(
R\kappa\right)  ^{2}$ taken in the dimensional units $\delta\nu\simeq
\delta\eta/T_{iat}$ strongly depends on the electron affinity energy and on
the condensate density via the dependence of $T_{iat}$ on these quantities.
Numerical values of $\delta\nu$ estimated for $Ca$ and $Cs$ are depicted in
Fig.\ref{Linewidth}, which shows that the width $\delta\nu$ is in the domain
of a typical diode laser linewidth for calcium already at $n\sim
10^{16}\mathrm{cm}^{-3}$.
Note that the tunneling time $\sim\mathcal{N}\sqrt{\eta}$ suggested by
$\sigma$-model (see Eq.(\ref{LinTProd}) with $q_{s}$ $\sim\sqrt{\eta}$ where
$\eta\rightarrow\eta-\sqrt{2}$) is not related with the inhomogeneous line
width by the uncertainty relations, and in the dimensional units it is of the
order of $T\sim\mathcal{N}\left(  R\kappa\right)  ^{-1}T_{iat}$ .

We now consider the frequency dependence of the number of atoms kicked out
from the condensate as the result of electron tunnel. Equation (\ref{Kick})
with the allowance of Eq.(\ref{GRANF}) for $3D$ Green's function can be
written in the form
\begin{align*}
\delta\mathcal{N}  &  =\frac{q_{s}N_{c}^{2}}{2\left(  \mathcal{N}\gamma
q_{s}\right)  ^{3}}\int G_{A}G_{R}\mathrm{d}\mathcal{V}\\
&  \sim\frac{q_{s}N_{c}^{2}}{2\left(  \mathcal{N}\gamma q_{s}\right)  ^{3}%
}q_{s}^{-d/2},
\end{align*}
which takes into account the number $N_{c}$ of the tunneling channels. One
sees that this dependence is a rather fast inverse power function of the This
form also shows that the condensate losses $\delta\mathcal{N}$ per electron
tunneling event decreases with increasing number of the condensate atoms. By
taking $q_{s}\sim1/R$, $N_{c}\sim\mathcal{N}/R\kappa$ one finds
\[
\delta\mathcal{N}\sim\frac{1}{\mathcal{N}\gamma^{3}}\left(  R\kappa\right)
^{d/2},
\]
and assuming $\gamma\sim1$ arrives at $\delta\mathcal{N}\sim\frac
{1}{\mathcal{N}}\left(  R\kappa\right)  ^{d/2}$. This number is of the order
of $10^{2}$ for $R\kappa\sim10^{4}$ and $\mathcal{N}\sim10^{6}$.

\section{Results and discussion\label{VI}}

Though our consideration seems to pose more questions than answer, we start
this section by emphasizing the novelty of the problem : in the most general
formulation, it addresses properties of a particle that experience an
essentially quantum multiple scattering off a material field in a
predetermined, in our case coherent, quantum state. The novelty has two
facets, one of which relates to the properties of the scattered particle that
in the presence of inelastic channels manifest some deviations from a similar
problem of quantum transport in disordered media, while the other facet
relates with possibility to reveal certain properties of an atomic condensate,
such as the size of fluctuations, by incorporating its quantum field to
electric circuits.

The main result of the paper is following. In spite of the fact that in Bose
condensate all atoms are in the same quantum state with a wave function
extending at microscopic distances, the tunneling electron feels the discrete
nature of individual atoms associated with the quantum density fluctuations,
and as a consequence, the tunneling gets strongly inhibited by the quantum
localization effect. In other words, a simple mean-field picture that
considers atomic condensate as a uniform support for the moving electron
cannot correspond to reality. Though the allowance for irreversible
transitions in the condensate partially destroys the electron interference and
thereby facilitate the tunneling, it does not remove the localization effect
completely. In the accordance with this picture is the fact that the state
density profile of the tunneling electron band has the Wigner semicircular
form in contrast to one corresponding to a free particle propagation.
Superposition of this profile with the profile of the localization length
determines the spectral line of the conductance efficiency.

The main message emerging from the consideration of relevant numbers is that
the electron tunneling through a condensate is not a completely impossible
process, and it can, at least in principle, be realized in a regime that
approaches the domain of parameters currently available experimentally. The
main problem is to identify atomic species that have weakly bounded negative
ions. The typically large ($\sim10^{3}A^{2}$) charge-exchange cross-section of
negative ions in normal conditions that can yet increase for cold collisions
are still not sufficient to ensure the efficient electron tunneling transport
at typical interatomic distances of currently existing condensates.
Nevertheless, since the charge exchange rate exponentially increases with
decreasing electron binding energy in negative ions, in some cases of closed
shells atoms with low polarizability, like $Ca$ for instance, the tunneling
regime can be achieved in principle, and the question is thus get moved to the
possibility to condense these atoms.

Note that under condition of the exponential dependence of the tunneling
probability on the interatomic distance, the role of careful calculation of
specific atomic parameters, such as the actual shape of wave function of the
attached electron, based on standard methods of atomic physics may becomes
important. It may change essentially the estimated required atomic density, as
it was the case for the presented calculations performed with and without the
parameter $\Lambda$. One should also have in mind the case of $He$ atoms with
closed $s$-shell that do not have electronic affinity at all since their
polarization potential is not strong enough to accommodate a bound state of
electron. Still they are able to create a disordered potential for the
tunneling electron at the energies close to zero. This potential will also
modify the state density profile at this region. All the calculations based on
$\sigma$-model thus remain valid, and the only difference will be in the sizes
of the time and energy units, that presumably should correspond to a more
efficient tunneling. Of course, one needs to know precise shape of the
electron binding potential of atoms and the corresponding scattering length in
order to make more accurate predictions.

Not a less important question is the efficiency of interface between the
electric leads and the condensate. In order to ensure an efficient tunneling
from the leads to the condensate without serious destruction of the latter,
one needs to have electronic surface states rather extended in the direction
perpendicular to the surface. Reliable determination of such states is a
subject of serious numerical work, but at the first glance one can hope that
the broad-band semiconductors may offer such a possibility. One also can
consider the possibility to employ carbon nanotubes for the purpose, having in
mind that the questions about the dielectric properties Ref.\cite{Kozinsky}
and Casimir potentials for the atom-tube interaction Ref.\cite{Knight} have
already been considered in details for such systems. However, for the case of
nanotubes one has to expect considerable decrease of the number of the
tunneling channels, which is a direct consequence of the small size of the
surface area.

It also worth mentioning that application of periodic external fields may not
only compensate for existing mismatches among the surface state band and the
condensate tunneling electron band, but also offer additional enhancement of
the transition between these bands when goes through highly excited and
therefore highly extended intermediate surface states. Moreover for a
multifrequency resonance setting, it can also serve as an efficient tool,
which reduces the influence of the thermal effects in the leads by satisfying
strict resonance conditions at the final or intermediate states of the
excitation cascade.

Looking at prospective, one can believe that the present consideration may
also become relevant in the context of condensed Rydberg atoms, where the
conductance might exist due to the partially occupied bands of delocalized
Rydberg states. However, in order to consider this case, one needs to allow
for the electronic many-body phenomena, such like Cooper pairs, that have been
omitted in the present work.

One should also have in mind that there exist other physical systems, such as
Bose condensates of excitons at the interface between two semiconductors,
where similar phenomena can take place.

\section{Acknowledgements}

V.M.A deeply appreciate numerous explanations of the details of nonlinear
$\sigma$-model technique given by Yan Fyodorov and Vladimir Kravtsov.
Finansial support of INTAS grant is greatly acknowledged by all the authors.

\section{Appendix\label{1}}

Here we show how to separate the atomic field operators from the ionic ones.
Prior to the separation, we make some transformations that simplify the
calculations. Consider Taylor expansions of the time dependent exponents in
Eq.(\ref{2.1}). Each term of the expansion is a sequence of the operator
products%
\begin{widetext}
\begin{equation}
\frac{\left(  -\mathrm{i}t\right)  ^{k}}{k!}\int\prod\limits_{i}^{k}%
\mathrm{d}r_{i}\ldots\hat{\Psi}^{+}(r_{3})\hat{\Phi}(r_{3})V(r_{3}%
-r_{2})\underbrace{\hat{\Phi}^{+}(r_{2})\hat{\Psi}(r_{2})\hat{\Psi}^{+}%
(r_{2})\hat{\Phi}(r_{2})}V(r_{2}-r_{1})\hat{\Phi}^{+}(r_{1})\hat{\Psi}(r_{1})
\label{ill}%
\end{equation}%
\end{widetext}
where the left argument of factor must coincide with the right argument of the
next operator, as it is emphasized in Eq.(\ref{ill}) for the second argument
by the underlining brace. This condition is the direct consequence of the
assumption that we have only one tunneling electron. The condensate operators
also correspond to the same coordinates, and hence they can be grouped
together when the Taylor series are summed back to the exponent. In other
words, operators $\hat{\Phi}^{+}(r)$ can be moved to $\hat{\Phi}(r)$
comprising the number of particles operator in each point where the ion field
operator $\hat{\Psi}$ is present. The same is valid for $\overline{\hat{\Psi}%
}$ operators. Equation (\ref{2.1}) thus takes the form%
\begin{widetext}%
\begin{align}
P(t,\tau,z)  &  =\left\langle 0\right\vert \left\langle 00\right\vert
_{i}\mathrm{e}^{-\mathrm{i}\int(A\hat{\Phi}^{+}(r)+A^{\ast}\hat{\Phi
}(r))\mathrm{d}r}\overline{\hat{\Psi}}(r_{a})\hat{\Phi}^{+}(r_{a})\hat{\Phi
}(r_{a})\mathrm{e}^{\mathrm{i}\tau\int\mathrm{d}r\mathrm{d}r_{1}\overline
{\hat{\Psi}}^{+}(r_{1})V(r-r_{1})\overline{\hat{\Psi}}(r)\hat{\Phi}^{+}%
(r)\hat{\Phi}(r)}\overline{\hat{\Psi}}^{+}(r_{b})\nonumber\\
&  \mathrm{e}^{\mathrm{i}\int(A\hat{\Phi}^{+}(r)+A^{\ast}\hat{\Phi
}(r))\mathrm{d}r}\mathrm{e}^{z\int\mathrm{d}r\hat{\Phi}^{+}(r)\hat{\Phi}%
(r)}\mathrm{e}^{-\mathrm{i}\int(A\hat{\Phi}^{+}(r)+A^{\ast}\hat{\Phi
}(r))\mathrm{d}r}\hat{\Phi}^{+}(r_{b})\hat{\Phi}(r_{b})\label{2.2}\\
&  \hat{\Psi}(r_{b})\mathrm{e}^{-\mathrm{i}t\int\mathrm{d}r\mathrm{d}r_{1}%
\hat{\Psi}^{+}(r_{1})V(r-r_{1})\hat{\Psi}(r)\hat{\Phi}^{+}(r)\hat{\Phi}%
(r)}\hat{\Psi}^{+}(r_{a})\mathrm{e}^{\mathrm{i}\int(A\hat{\Phi}^{+}%
(r)+A^{\ast}\hat{\Phi}(r))\mathrm{d}r}\left\vert 00\right\rangle
_{i}\left\vert 0\right\rangle .\nonumber
\end{align}%
\end{widetext}%

The general idea underlying the separation of the ionic and condensate fields
rely on functional integrals, and in particular, on the functional Fourier
transformation. The main steps of this procedure are as follows. A functional
$\Omega\left(  \widehat{X}(x)\widehat{Y}(x)\right)  $ depending on a product
of two commuting operator fields $\widehat{X}(x)$ and $\widehat{Y}(x)$ can be
written as a functional integral%
\[
\int\mathcal{D}u(x)\Omega\left(  \widehat{X}(x)u(x)\right)  \delta_{F}\left(
\widehat{Y}(x)-u(x)\right)
\]
where the functional Dirac $\delta$-function has the Fourier representation
\[
\delta_{F}\left(  f(x)-g(x)\right)  =\int\mathrm{e}^{\mathrm{i}2\pi
\int\mathrm{d}x\left[  f(x)-g(x)\right]  u^{\ast}(x)}\mathcal{D}u^{\ast}(x).
\]
Here $u(x)\mathcal{\ }$and $u^{\ast}(x)$ \ are two independent real-valued
functional variables. By combining these two expressions one represents
$\Omega\left(  \widehat{X}(x)\widehat{Y}(x)\right)  $ as a functional integral
of a product, where one factor depends only on the operator field $\widehat
{X}(x)$ and the other only on the operator field $\widehat{Y}(x)$ :
\[
\int\Omega\left(  \widehat{X}(x)u(x)\right)  \mathrm{e}^{\mathrm{i}2\pi
\int\mathrm{d}x\left[  \widehat{Y}(x)-u(x)\right]  u^{\ast}(x)}\mathcal{D}%
u(x)\mathcal{D}u^{\ast}(x).
\]
In particular, two factors in Eq.(\ref{2.2}),
\[
\mathrm{e}^{\mathrm{i}\tau\int\mathrm{d}r\mathrm{d}r_{1}\overline{\hat{\Psi}%
}^{+}(r_{1})V(r-r_{1})\overline{\hat{\Psi}}(r)\hat{\Phi}^{+}(r)\hat{\Phi}(r)}%
\]
and%
\[
\mathrm{e}^{-\mathrm{i}t\int\mathrm{d}r\mathrm{d}r_{1}\hat{\Psi}^{+}%
(r_{1})V(r-r_{1})\hat{\Psi}(r)\hat{\Phi}^{+}(r)\hat{\Phi}(r)}%
\]
can be represented in the form
\begin{align*}
&  \int\mathrm{e}^{\mathrm{i}\tau\int\mathrm{d}r\mathrm{d}r_{1}\overline
{\hat{\Psi}}^{+}(r_{1})V(r-r_{1})\overline{\hat{\Psi}}(r)u(r)}\\
&  \mathrm{e}^{\mathrm{i}2\pi\int\mathrm{d}r\left[  \hat{\Phi}^{+}(r)\hat
{\Phi}(r)-u(r)\right]  u^{\ast}(r)}\mathcal{D}u(x)\mathcal{D}u^{\ast}(x)
\end{align*}
and%
\begin{align*}
&  \int\mathrm{e}^{-\mathrm{i}t\int\mathrm{d}r\mathrm{d}r_{1}\hat{\Psi}%
^{+}(r_{1})V(r-r_{1})\hat{\Psi}(r)v(r)}\\
&  \mathrm{e}^{\mathrm{i}2\pi\int\mathrm{d}r\left[  \hat{\Phi}^{+}(r)\hat
{\Phi}(r)-v(r)\right]  v^{\ast}(r)}\mathcal{D}v(x)\mathcal{D}v^{\ast}(x),
\end{align*}
respectively.

By regrouping the operators, one obtains Eq.(\ref{2.2}) as a functional
integral of a product of three matrix elements each of which is taken by the
corresponding vacuum state%
\begin{align}
P(t,\tau,z)  &  =\int\mathcal{D}v(x)\mathcal{D}v^{\ast}(x)\mathcal{D}%
u(x)\mathcal{D}u^{\ast}(x)\label{2.2a}\\
&  \mathrm{e}^{-\mathrm{i}2\pi v(r)v^{\ast}(r)}\mathrm{e}^{-\mathrm{i}2\pi
u(r)u^{\ast}(r)}\left\langle Cd\right\rangle G_{A}\left(  \tau\right)
G_{R}\left(  t\right)  ,\nonumber
\end{align}
where
\begin{align}
&  \overline{G}_{A}=\left\langle 0\right\vert \overline{\hat{\Psi}}%
(r_{a})\mathrm{e}^{\mathrm{i}\tau\int\mathrm{d}r\mathrm{d}r_{1}\overline
{\hat{\Psi}}^{+}(r_{1})V(r-r_{1})\overline{\hat{\Psi}}(r)u(r)}\overline
{\hat{\Psi}}^{+}(r_{b})\left\vert 0\right\rangle \nonumber\\
&  \overline{G}_{R}=\left\langle 0\right\vert \hat{\Psi}(r_{b})\mathrm{e}%
^{-\mathrm{i}t\int\mathrm{d}r\mathrm{d}r_{1}\hat{\Psi}^{+}(r_{1}%
)V(r-r_{1})\hat{\Psi}(r)v(r)}\hat{\Psi}^{+}(r_{a})\left\vert 0\right\rangle
\nonumber
\end{align}
are the advanced and retarded Green's functions calculated over the vacuum
states of the ion, and the matrix element over the condensate vacuum%
\begin{align*}
\left\langle Cd\right\rangle  &  =\left\langle 0\right\vert \mathrm{e}%
^{-\mathrm{i}\int(A\hat{\Phi}^{+}(r)+A^{\ast}\hat{\Phi}(r))\mathrm{d}r}%
\hat{\Phi}^{+}(r_{a})\hat{\Phi}(r_{a})\\
&  \mathrm{e}^{\mathrm{i}2\pi\int\mathrm{d}r\hat{\Phi}^{+}(r)\hat{\Phi
}(r)u^{\ast}(r)}\mathrm{e}^{\mathrm{i}\int(A\hat{\Phi}^{+}(r)+A^{\ast}%
\hat{\Phi}(r))\mathrm{d}r}\\
&  \mathrm{e}^{z\int\mathrm{d}r\hat{\Phi}^{+}(r)\hat{\Phi}(r)}\mathrm{e}%
^{-\mathrm{i}\int(A\hat{\Phi}^{+}(r)+A^{\ast}\hat{\Phi}(r))\mathrm{d}r}%
\hat{\Phi}^{+}(r_{b})\hat{\Phi}(r_{b})\\
&  \mathrm{e}^{\mathrm{i}2\pi\int\mathrm{d}r\hat{\Phi}^{+}(r)\hat{\Phi
}(r)v^{\ast}(r)}\mathrm{e}^{\mathrm{i}\int(A\hat{\Phi}^{+}(r)+A^{\ast}%
\hat{\Phi}(r))\mathrm{d}r}\left\vert 0\right\rangle .
\end{align*}
The bar over the Green's functions $\overline{G}_{A}$ and $\overline{G}_{R}$
shows that they are calculated for the operators that include fields $u(r)$
and $v(r)$, respectively.

We calculate $\left\langle Cd\right\rangle $ with the help of two types of
operator relations : the Baker-Campbell-Hausdorff rules%
\begin{align}
\mathrm{e}^{A+B}  &  =\mathrm{e}^{B}\mathrm{e}^{A}\mathrm{e}^{\left[
A,B\right]  /2}\label{BCH}\\
\mathrm{e}^{A}\mathrm{e}^{B}  &  =\mathrm{e}^{B}\mathrm{e}^{A}\mathrm{e}%
^{\left[  A,B\right]  }\nonumber
\end{align}
for the operators with c-number commutator, and the relations
\begin{align}
\mathrm{e}^{z\int\mathrm{d}r\hat{\Phi}^{+}(r)\hat{\Phi}(r)}\hat{\Phi}%
^{+}(r)\hat{\Phi}(r)\mathrm{e}^{-z\int\mathrm{d}r\hat{\Phi}^{+}(r)\hat{\Phi
}(r)}  &  =\mathrm{e}^{z}\hat{\Phi}^{+}(r)\nonumber\\
\mathrm{e}^{z\int\mathrm{d}r\hat{\Phi}^{+}(r)\hat{\Phi}(r)}\hat{\Phi}%
^{+}(r)\hat{\Phi}(r)\mathrm{e}^{-z\int\mathrm{d}r\hat{\Phi}^{+}(r)\hat{\Phi
}(r)}  &  =\mathrm{e}^{-z}\hat{\Phi}(r) \label{Osc}%
\end{align}
for the bosonic operators. With the help of Eq.(\ref{Osc}) one can move the
factors depending on the number operators towards the right vacuum state,
where they just result in factors $1$, and obtain%
\begin{align*}
&  \left\langle 0\right\vert \mathrm{e}^{-\mathrm{i}\int(A\mathrm{e}%
^{-\mathrm{i}2\pi u^{\ast}(r)}\hat{\Phi}^{+}(r)+A^{\ast}\mathrm{e}%
^{\mathrm{i}2\pi u^{\ast}(r)}\hat{\Phi}(r))\mathrm{d}r}\hat{\Phi}^{+}%
(r_{a})\hat{\Phi}(r_{a})\\
&  \mathrm{e}^{\mathrm{i}\int(A\hat{\Phi}^{+}(r)+A^{\ast}\hat{\Phi
}(r))\mathrm{d}r}\mathrm{e}^{-\mathrm{i}\int(A\mathrm{e}^{z}\hat{\Phi}%
^{+}(r)+A^{\ast}\mathrm{e}^{-z}\hat{\Phi}(r))\mathrm{d}r}\\
&  \hat{\Phi}^{+}(r_{b})\hat{\Phi}(r_{b})\mathrm{e}^{\mathrm{i}\int
(A\mathrm{e}^{z+\mathrm{i}2\pi v^{\ast}(r)}\hat{\Phi}^{+}(r)+A^{\ast
}\mathrm{e}^{-z-\mathrm{i}2\pi v^{\ast}(r)}\hat{\Phi}(r))\mathrm{d}%
r}\left\vert 0\right\rangle .
\end{align*}
Ignoring the quantum fluctuations in the pre-exponential factors yields%
\begin{align*}
\left\langle Cd\right\rangle =  &  n^{2}\left\langle 0\right\vert
\mathrm{e}^{-\mathrm{i}\int(A\mathrm{e}^{-\mathrm{i}2\pi u^{\ast}(r)}\hat
{\Phi}^{+}(r)+A^{\ast}\mathrm{e}^{\mathrm{i}2\pi u^{\ast}(r)}\hat{\Phi
}(r))\mathrm{d}r}\\
&  \mathrm{e}^{\mathrm{i}\int(A\hat{\Phi}^{+}(r)+A^{\ast}\hat{\Phi
}(r))\mathrm{d}r}\mathrm{e}^{-\mathrm{i}\int(A\mathrm{e}^{z}\hat{\Phi}%
^{+}(r)+A^{\ast}\mathrm{e}^{-z}\hat{\Phi}(r))\mathrm{d}r}\\
&  \mathrm{e}^{\mathrm{i}\int(A\mathrm{e}^{z+\mathrm{i}2\pi v^{\ast}(r)}%
\hat{\Phi}^{+}(r)+A^{\ast}\mathrm{e}^{-z-\mathrm{i}2\pi v^{\ast}(r)}\hat{\Phi
}(r))\mathrm{d}r}\left\vert 0\right\rangle .
\end{align*}
With the help of Eq.(\ref{BCH}) the exponential terms can be normally ordered,
that is rearranged in such a way that all $\hat{\Phi}$ are to the right with
respect to $\hat{\Phi}^{+}$. All the terms containing normally ordered field
operators vanish. The c-number commutators appearing in the course of the
normal ordering yield a factor $\mathrm{e}^{w(u^{\ast}(r),v^{\ast}(r))}$,
which reads%
\begin{align*}
w(u^{\ast}(r),v^{\ast}(r))  &  =n\int\mathrm{d}r(\mathrm{e}^{\mathrm{i}2\pi
u^{\ast}(r)}+\mathrm{e}^{\mathrm{i}2\pi v^{\ast}(r)}-2+\\
&  \mathrm{e}^{z}(1-\mathrm{e}^{\mathrm{i}2\pi u^{\ast}(r)})(1-\mathrm{e}%
^{\mathrm{i}2\pi v^{\ast}(r)})).
\end{align*}

The functional Fourier transform
\begin{equation}
W_{u,v}=\int\mathcal{D}v^{\ast}(x)\mathcal{D}u^{\ast}(x)\mathrm{e}%
^{w-2\pi\mathrm{i}\int\mathrm{d}r\left[  u(r)u^{\ast}(r)+(r)v^{\ast
}(r)\right]  }, \label{FFT}%
\end{equation}
of $\mathrm{e}^{w(u^{\ast}(r),v^{\ast}(r))}$ plays role of the weight factor
for functional integration over the fields $u(r)$ and $v(r)$. However, it has
a structure inconvenient for calculations. Technically is much simpler to
perform the functional integration over the fields $u$ and $v$ \ first, and
than integrate over $u^{\ast}(r)$ and $v^{\ast}(r)$.

Equation (\ref{2.2a}) thus takes the form%
\begin{align*}
P(t,\tau,z)  &  =n^{2}\int\mathcal{D}v^{\ast}(x)\mathcal{D}u^{\ast
}(x)\mathcal{D}v(x)\mathcal{D}u(x)\\
&  \overline{G}_{A}\left(  \tau\right)  \overline{G}_{R}\left(  t\right)
\mathrm{e}^{w-2\pi\mathrm{i}\int\mathrm{d}r\left[  u(r)u^{\ast}(r)+(r)v^{\ast
}(r)\right]  },
\end{align*}
One can perform the Fourier transformation over times $t$ and $\tau$ and write%
\begin{align}
P(\varepsilon,\xi,z)  &  =n^{2}\int W_{u,v}\mathcal{D}v(x)\mathcal{D}%
u(x)\label{Prob}\\
&  \overline{G}_{A}\left(  \xi\right)  \overline{G}_{R}\left(  \varepsilon
\right)  \mathrm{e}^{w-2\pi\mathrm{i}\int\mathrm{d}r\left[  u(r)u^{\ast
}(r)+(r)v^{\ast}(r)\right]  .}\nonumber
\end{align}
where the Green's functions can be given explicitly%
\begin{align}
\overline{G}_{A}\left(  \xi\right)   &  =\left\langle 0\right\vert
\overline{\hat{\Psi}}(r_{a})\left(  \xi-\widehat{H}_{u}\right)  ^{-1}%
\overline{\hat{\Psi}}^{+}(r_{b})\left\vert 0\right\rangle \label{A.1}\\
\overline{G}_{R}\left(  \varepsilon\right)   &  =\left\langle 0\right\vert
\hat{\Psi}(r_{b})\left(  \varepsilon-\widehat{H}_{v}\right)  ^{-1}\hat{\Psi
}^{+}(r_{a})\left\vert 0\right\rangle \nonumber
\end{align}
in terms of the field operators%
\begin{align*}
\widehat{H}_{u}  &  =\int\mathrm{d}r\mathrm{d}r_{1}\overline{\hat{\Psi}}%
^{+}(r_{1})V(r-r_{1})\overline{\hat{\Psi}}(r)u(r)\\
\widehat{H}_{v}  &  =\int\mathrm{d}r\mathrm{d}r_{1}\hat{\Psi}^{+}%
(r_{1})V(r-r_{1})\hat{\Psi}(r)v(r),
\end{align*}
that differ from the Hamiltonians Eq.(\ref{HAHR})\ having classical fields
$u(r)$ and $v(r)$ instead of the quantum condensate fields $\hat{\Phi}^{+}(r)$
and $\hat{\Phi}(r)$ .

In order to find the average of $\overline{G}_{A}\left(  \xi\right)
\overline{G}_{R}\left(  \varepsilon\right)  $, we follow the prescription well
elaborated in the theory of disordered systems and express the Green's
functions with the help of functional superintegrals over supervector fields
$\lambda(r)$ and $\mu(r)$. For the advanced and retarded Green's functions
these fields are grouped to the line vectors%
\[
\lambda\left(  r\right)  =\left(
\begin{array}
[c]{cc}%
\lambda_{b}\left(  r\right)  & \lambda_{f}\left(  r\right)
\end{array}
\right)  ;~\mu\left(  r\right)  =\left(
\begin{array}
[c]{cc}%
\mu_{b}\left(  r\right)  & \mu_{f}\left(  r\right)
\end{array}
\right)
\]
respectively, and the conjugated columns vectors%
\[
\overline{\lambda}\left(  r\right)  =\left(
\begin{array}
[c]{c}%
\overline{\lambda}_{b}\left(  r\right) \\
\overline{\lambda}_{f}\left(  r\right)
\end{array}
\right)  ;~\overline{\mu}_{b}\left(  r\right)  =\left(
\begin{array}
[c]{c}%
\overline{\mu}_{b}\left(  r\right) \\
\overline{\mu}_{f}\left(  r\right)
\end{array}
\right)  .
\]
This integral over regular(bosonic) functional variables $\lambda_{b}(r)$ and
$\mu_{b}(r)$ and anticommuting grassmannian (fermionic) variables $\lambda
_{f}(r)$ and $\mu_{f}(r)$ read
\begin{align}
\overline{G}_{R}\left(  \varepsilon\right)   &  =\int\mathcal{D}%
\lambda(r)\mathcal{D}\overline{\lambda}(r)\mathrm{e}^{\mathrm{i}%
S_{\varepsilon}}\overline{\lambda}_{f}(r_{a})\lambda_{f}(r_{b})\nonumber\\
\overline{G}_{A}\left(  \xi\right)   &  =\int\mathcal{D}\mu(r)\mathcal{D}%
\overline{\mu}(r)\mathrm{e}^{-\mathrm{i}S_{\xi})}\mu_{f}(r_{b})\overline{\mu
}_{f}(r_{a}), \label{Green}%
\end{align}
where the conjugated variables denoted by bars and the actions have the form%
\begin{align}
S_{\varepsilon}  &  =\int\mathrm{d}r\lambda(r)\varepsilon\overline{\lambda
}(r)-\int\mathrm{d}r\mathrm{d}r_{1}\lambda(r_{1})\widehat{V}(r-r_{1}%
)\overline{\lambda}(r)v(r)\nonumber\\
S_{\xi}  &  =\int\mathrm{d}r\mu(r)\xi\overline{\mu}(r)-\int\mathrm{d}%
r\mathrm{d}r_{1}\mu(r_{1})\widehat{V}(r-r_{1})\overline{\mu}(r)u(r).
\label{Actions}%
\end{align}
Here the products $\lambda(r)\overline{\lambda}(r)$ and $\mu(r)\overline{\mu
}(r)$ stand for the scalar products $\lambda_{b}(r)\overline{\lambda}%
_{b}(r)+\lambda_{f}(r)\overline{\lambda}_{f}(r)$ and $\mu_{b}(r)\overline{\mu
}_{b}(r)+\mu_{f}(r)\overline{\mu}_{f}(r)$, respectively. The functional
differentials $\mathcal{D}\lambda(r)\mathcal{D}\overline{\lambda}(r)$ stand
for $\mathcal{D}\lambda_{b}(r)\mathcal{D}\overline{\lambda}_{b}(r)\mathcal{D}%
\lambda_{f}(r)\mathcal{D}\overline{\lambda}_{f}(r)$, and $\mathcal{D}%
\mu(r)\mathcal{D}\overline{\mu}(r)$ for $\mathcal{D}\mu_{b}(r)\mathcal{D}%
\overline{\mu}_{b}(r)\mathcal{D}\mu_{f}(r)\mathcal{D}\overline{\mu}_{f}(r)$,
and the operator $V$ reads%

\[
\widehat{V}(r-r_{1})=\left(
\begin{array}
[c]{cc}%
V(r-r_{1}) & 0\\
0 & V(r-r_{1})
\end{array}
\right)  .
\]

We substitute Eq.(\ref{Green}) with the actions Eq.(\ref{Actions}) to
Eq.(\ref{Prob}) and perform the integration over the fields $u$ and $v$ first.
\ This integration yields $\delta_{F}$ functions with the arguments $2\pi
v^{\ast}(x)-\int\mathrm{d}r_{1}\lambda(r_{1})\widehat{V}(r-r_{1}%
)\overline{\lambda}(r)$ and $2\pi u^{\ast}(x)+\int\mathrm{d}r_{1}\mu
(r_{1})\widehat{V}(r-r_{1})\overline{\mu}(r)$. Performing the integration over
the variables $u^{\ast}$ and $v^{\ast}$ one finds%
\begin{widetext}%
\begin{align}
P(\xi,\varepsilon,z)=  &  n^{2}\int\mathcal{D}\lambda(r)\mathcal{D}%
\overline{\lambda}(r)\mathrm{e}^{\mathrm{i}\int\mathrm{d}r\lambda
(r)\varepsilon\overline{\lambda}(r)}\overline{\lambda}_{f}(r_{a})\lambda
(r_{b})\int\mathcal{D}\mu(r)\mathcal{D}\overline{\mu}(r)\mathrm{e}%
^{-\mathrm{i}\int\mathrm{d}r\mu(r)\xi\overline{\mu}(r)}\mu_{f}(r_{b}%
)\overline{\mu}_{f}(r_{a})\nonumber\\
&  \exp\left\{  n\int\mathrm{d}r(\mathrm{e}^{\mathrm{i}\int\mathrm{d}%
r_{1}\lambda(r_{1})V(r-r_{1})\overline{\lambda}(r)}+\mathrm{e}^{-\mathrm{i}%
\int\mathrm{d}r_{1}\mu(r_{1})V(r-r_{1})\overline{\mu}(r)}-2)\right\}
\label{ProbInt}\\
&  \exp\left\{  n\int\mathrm{d}r(1-\mathrm{e}^{\mathrm{i}\int\mathrm{d}%
r_{1}\lambda(r_{1})V(r-r_{1})\overline{\lambda}(r)})(1-\mathrm{e}%
^{-\mathrm{i}\int\mathrm{d}r_{1}\mu(r_{1})V(r-r_{1})\overline{\mu}%
(r)})\mathrm{e}^{z}\right\} \nonumber
\end{align}%
\end{widetext}%

Thus far the calculations were exact. At this point one has to make an
approximation, since exact calculation of the functional integrals are further
impossible. \ We notice that one fold integrals with the exponentially
decreasing kernel $V(r-r_{1})$ are of the order on the atomic density $n$,
which is a small dimensionless parameter. By the second order Taylor expansion
of the terms entering the exponents in the second and third lines of
Eq.(\ref{ProbInt}) we obtain%
\begin{widetext}
\begin{align}
P(\xi,\varepsilon,z)=  &  n^{2}\int\mathcal{D}\lambda(r)\mathcal{D}%
\overline{\lambda}(r)\mathcal{D}\mu(r)\mathcal{D}\overline{\mu}(r)\overline
{\lambda}_{f}(r_{a})\lambda(r_{b})\mu_{f}(r_{b})\overline{\mu}_{f}%
(r_{a})\mathrm{e}^{\mathrm{i}\int\mathrm{d}r\lambda(r)\varepsilon
\overline{\lambda}(r)-\mathrm{i}\int\mathrm{d}r\mu(r)\xi\overline{\mu}%
(r)}\nonumber\\
&  \mathrm{e}^{\mathrm{i}n\int\left[  \lambda(r_{1})V(r-r_{1})\overline
{\lambda}(r)-\mu(r_{1})V(r-r_{1})\overline{\mu}(r)\right]  \mathrm{d}%
r_{1}\mathrm{d}r+\mathrm{e}^{z}n\int\mathrm{d}r_{1}\mathrm{d}r_{2}%
\mathrm{d}r{\Huge [}\lambda(r_{1})V(r-r_{1})\overline{\lambda}(r)\mu
(r_{2})V(r-r_{2})\overline{\mu}(r){\Huge ]}}\nonumber\\
&  \mathrm{e}^{-\frac{n}{2}\int\mathrm{d}r_{1}\mathrm{d}r_{2}\mathrm{d}%
r{\Huge [}\lambda(r_{1})V(r-r_{1})\overline{\lambda}(r)\lambda(r_{2}%
)V(r-r_{2})\overline{\lambda}(r)+\mu(r_{1})V(r-r_{1})\overline{\mu}%
(r)\mu(r_{2})V(r-r_{2})\overline{\mu}(r){\Huge ]}} \label{ProbTay}%
\end{align}%
\end{widetext}
Let us rewrite this expression in a more compact way by introducing two
supervectors in the functional coordinate space -\ the column vector
$\left\vert \overline{\phi}\right\rangle $ defined by their coordinate
representations as
\[
\overline{\phi}\left(  r\right)  =\left\langle r\right\vert \left.
\overline{\phi}\right\rangle =\left(
\begin{array}
[c]{c}%
\overline{\lambda}_{b}\left(  r\right) \\
\overline{\mu}_{b}\left(  r\right) \\
\overline{\lambda}_{f}\left(  r\right) \\
\overline{\mu}_{f}\left(  r\right)
\end{array}
\right)
\]
and the corresponding line vector $\left\langle \phi\right\vert $%
\[
\phi\left(  r^{\prime}\right)  =\left\langle \phi\right\vert \left.
r^{\prime}\right\rangle =\left(
\begin{array}
[c]{cccc}%
\lambda_{b}\left(  r^{\prime}\right)  & \mu_{b}\left(  r^{\prime}\right)  &
\lambda_{f}\left(  r^{\prime}\right)  & \mu_{f}\left(  r^{\prime}\right)
\end{array}
\right)  .
\]
In this notations Eq.(\ref{ProbTay}) takes the form%
\begin{widetext}
\begin{align}
P(\xi,\varepsilon,z)=  &  n^{2}\int\mathcal{D}\left\vert \phi\right\rangle
\mathcal{D}\left\langle \overline{\phi}\right\vert ~\overline{\lambda}%
_{f}(r_{a})\lambda(r_{b})\mu_{f}(r_{b})\overline{\mu}_{f}(r_{a})\mathrm{\exp
}\left[  \left\langle \phi\right\vert \left(  \mathrm{i}\mathbf{\zeta
+\mathrm{i}}\eta\mathbf{\mathbf{\Sigma}_{z}+}\mathrm{i}n\mathbf{\mathbf{\Sigma
}_{z}V}\right)  \left\vert \overline{\phi}\right\rangle \right] \label{ProbM}%
\\
&  \mathrm{\exp}\left[  -\frac{n}{2}(1-\mathrm{e}^{z})\int\mathrm{d}%
r_{1}\mathrm{d}r_{2}\mathrm{d}r\left\langle \phi\left(  r_{2}\right)
V(r_{2}-r)\mathbf{\Sigma}_{z}\overline{\phi}\left(  r\right)  \right\rangle
\left\langle \phi\left(  r_{1}\right)  V(r_{1}-r)\mathbf{\Sigma}_{z}%
\overline{\phi}\left(  r\right)  \right\rangle \right] \nonumber\\
&  \mathrm{\exp}\left[  -\frac{n}{2}(1+\mathrm{e}^{z})\int\mathrm{d}%
r_{1}\mathrm{d}r_{2}\mathrm{d}r\left\langle \phi\left(  r_{2}\right)
V(r-r_{2})\overline{\phi}\left(  r\right)  \right\rangle \left\langle
\phi\left(  r_{1}\right)  V(r-r_{1})\overline{\phi}\left(  r\right)
\right\rangle \right]  ,\nonumber
\end{align}
where two matrices%
\[
\left\langle r^{\prime}\right\vert \mathbf{V}\left\vert r\right\rangle
=V(r^{\prime}-r)\left(
\begin{array}
[c]{cccc}%
1 & 0 & 0 & 0\\
0 & 1 & 0 & 0\\
0 & 0 & 1 & 0\\
0 & 0 & 0 & 1
\end{array}
\right)  ~\mathrm{and}~\left\langle r^{\prime}\right\vert \mathbf{\Sigma}%
_{z}\left\vert r\right\rangle =\delta(r^{\prime}-r)\widehat{\Sigma}_{z}%
=\delta(r^{\prime}-r)\left(
\begin{array}
[c]{cccc}%
1 & 0 & 0 & 0\\
0 & -1 & 0 & 0\\
0 & 0 & 1 & 0\\
0 & 0 & 0 & -1
\end{array}
\right)
\]%
\end{widetext}
and two variables $\eta=\frac{\varepsilon+\xi}{2}$ and $\zeta=\frac
{\varepsilon-\xi}{2}$ have been introduced. Here by $\mathcal{D}\left\vert
\phi\right\rangle \mathcal{D}\left\langle \overline{\phi}\right\vert $ we
denote the functional integration by all eight components of the vectors
$\left\vert \phi\right\rangle \mathcal{\ }$and $\left\langle \overline{\phi
}\right\vert $.

Let us rewrite Eq.(\ref{ProbM}) in an even shorter way by introducing an
object similar to the density matrix -- the direct product%
\[
\mathbf{\rho}=\left\vert \overline{\phi}\right\rangle \otimes\left\langle
\phi\right\vert .
\]
This allows one to put Eq.(\ref{ProbTay}) in the form%
\begin{align}
P(\zeta,\eta,z)=  &  n^{2}\int\mathcal{D}\left\vert \overline{\phi
}\right\rangle \mathcal{D}\left\langle \phi\right\vert ~\overline{\lambda}%
_{f}(r_{a})\lambda(r_{b})\mu_{f}(r_{b})\overline{\mu}_{f}(r_{a})\nonumber\\
&  \mathrm{e}^{\mathrm{STr}\left(  \mathrm{i}\mathbf{\zeta+\mathrm{i}}%
\eta\mathbf{\mathbf{\Sigma}_{z}+}\mathrm{i}n\mathbf{\mathbf{\Sigma}_{z}%
V}\right)  \mathbf{\rho}}\label{ProbStr}\\
&  \mathrm{e}^{-\frac{n}{4}(1-\mathrm{e}^{z})\mathrm{STr}\mathbf{\rho V\Sigma
}_{z}\mathbf{\rho V\Sigma}_{z}-\frac{n}{4}(1+\mathrm{e}^{z})\mathrm{STr}%
\mathbf{\rho V\rho V}}.\nonumber
\end{align}
The symbol $\mathrm{STr}$ stands for the regular trace operator with respect
to the coordinate dependence and with respect to regular vector components and
implies change of the sign for the Grassmann components. By $\mathrm{Str}$ we
denote the supertrace operation, which involves only the matrix variables and
does not imply integration over the coordinates, it amounts to the difference
of the traces of the boson-boson and fermion-fermion blocks.

The next step is application of the Hubbard-Stratanovich separation of the
terms quadratic in $\mathbf{\rho}$. This can be done by introducing an
superintegral over a supermatrix $\mathbf{Q}$ which, generally speaking
depends only the coordinate $r$, that is $\left\langle r^{\prime}\right\vert
\mathbf{Q}\left\vert r\right\rangle =\delta(r^{\prime}-r)\widehat{Q}$ with
\begin{equation}
\widehat{Q}=\left(
\begin{array}
[c]{cccc}%
Q_{\lambda_{b}\lambda_{b}} & Q_{\lambda_{b}\mu_{b}} & Q_{\lambda_{b}%
\lambda_{f}} & Q_{\lambda_{b}\mu_{f}}\\
Q_{\mu_{b}\lambda_{b}} & Q_{\mu_{b}\mu_{b}} & Q_{\mu_{b}\lambda_{f}} &
Q_{\mu_{b}\mu_{f}}\\
Q_{\lambda_{f}\lambda_{b}} & Q_{\lambda_{f}\mu_{b}} & Q_{\lambda_{f}%
\lambda_{f}} & Q_{\lambda_{f}\mu_{f}}\\
Q_{\mu_{f}\lambda_{b}} & Q_{\mu_{f}\mu_{b}} & Q_{\mu_{f}\lambda_{f}} &
Q_{\mu_{f}\mu_{f}}%
\end{array}
\right)  , \label{Q}%
\end{equation}
where the matrix elements $Q_{i,j}=Q_{i,j}(r)$ are the regular functions of
$r$ in the diagonal $2\times2$ blocks (boson-boson and fermion-fermion) \ and
the Grassmann functions of $r$ in the off-diagonal (boson-fermion) blocks. The
third line of Eq.(\ref{ProbStr}) then takes the form%
\begin{widetext}%
\begin{equation}
\mathrm{\exp}\left\{  -\frac{n}{4}\mathrm{STr}\left[  (1-\mathrm{e}%
^{z})\mathbf{\rho V\Sigma}_{z}\mathbf{\rho V\Sigma}_{z}+(1+\mathrm{e}%
^{z})\mathbf{\rho V\rho V}\right]  \right\}  =\int\mathcal{D}\mathbf{Q}%
(r)\mathrm{\exp}\left[  -\frac{n}{2}\mathrm{STr}\mathbf{Q}^{2}-\mathrm{i}%
~n\mathrm{STr}\mathbf{VQ}^{\prime}\mathbf{\rho}\right]  \label{HS}%
\end{equation}
where $\left\langle r^{\prime}\right\vert \mathbf{Q}^{\prime}\left\vert
r\right\rangle =\delta(r^{\prime}-r)\widehat{Q}^{\prime}$ with
\[
\widehat{Q}^{\prime}=\left(
\begin{array}
[c]{cccc}%
-Q_{\lambda_{b}\lambda_{b}} & \mathrm{ie}^{z/2}Q_{\lambda_{b}\mu_{b}} &
-Q_{\lambda_{b}\lambda_{f}} & \mathrm{ie}^{z/2}Q_{\lambda_{b}\mu_{f}}\\
\mathrm{ie}^{z/2}Q_{\mu_{b}\lambda_{b}} & Q_{\mu_{b}\mu_{b}} & \mathrm{ie}%
^{z/2}Q_{\mu_{b}\lambda_{f}} & Q_{\mu_{b}\mu_{f}}\\
-Q_{\lambda_{f}\lambda_{b}} & \mathrm{ie}^{z/2}Q_{\lambda_{f}\mu_{b}} &
-Q_{\lambda_{f}\lambda_{f}} & \mathrm{ie}^{z/2}Q_{\lambda_{f}\mu_{f}}\\
\mathrm{ie}^{z/2}Q_{\mu_{f}\lambda_{b}} & Q_{\mu_{f}\mu_{b}} & \mathrm{ie}%
^{z/2}Q_{\mu_{f}\lambda_{f}} & Q_{\mu_{f}\mu_{f}}%
\end{array}
\right)  .
\]%
\end{widetext}
In these notations Eq.(\ref{ProbStr}) reads%
\begin{align}
&  P(\zeta,\eta,z)=n^{2}\int\mathcal{D}\left\vert \overline{\phi}\right\rangle
\mathcal{D}\left\langle \phi\right\vert \mathcal{D}\mathbf{Q}(r)\mathrm{e}%
^{-\frac{n}{2}\mathrm{STr}\mathbf{Q}^{2}}\nonumber\\
&  ~\overline{\lambda}_{f}(r_{a})\lambda(r_{b})\mu_{f}(r_{b})\overline{\mu
}_{f}(r_{a})\mathrm{e}^{\mathrm{STr}\left(  \mathrm{i}\mathbf{\zeta
+\mathrm{i}}\eta\mathbf{\mathbf{\Sigma}_{z}+}\mathrm{i}n\mathbf{\mathbf{\Sigma
}_{z}V-}\mathrm{i}~n\mathbf{VQ}^{\prime}\right)  \mathbf{\rho}}.
\label{ProbDQ}%
\end{align}
We note another possibility, - the factors $\mathrm{e}^{z/2}$ can be moved
from the matrix $\widehat{Q}^{\prime}$ to the matrix $\widehat{Q}$ . This
transformation has a unit Berezian (the supermatrix analog of Jacobian) since
factors $\mathrm{e}^{-z/2}$ and $\mathrm{e}^{z/2}$ arising from four bosonic
and four fermionic variables cancel each other, and Eq.(\ref{ProbDQ}) takes
the form
\begin{align}
&  P(\zeta,\eta,z)=n^{2}\int\mathcal{D}\left\vert \overline{\phi}\right\rangle
\mathcal{D}\left\langle \phi\right\vert \mathcal{D}\mathbf{Q}(r)\overline
{\lambda}_{f}(r_{a})\lambda_{f}(r_{b})\label{ProbDQ1}\\
&  \mu_{f}(r_{b})\overline{\mu}_{f}(r_{a})\mathrm{e}^{\mathrm{STr}\left(
\mathrm{i}\mathbf{\zeta+\mathrm{i}}\eta\mathbf{\mathbf{\Sigma}_{z}+}%
\mathrm{i}n\mathbf{\mathbf{\Sigma}_{z}V+}\mathrm{i}~n\mathbf{V}\sqrt
{\mathbf{\mathbf{\Sigma}_{z}}}\mathbf{Q}\sqrt{\mathbf{\mathbf{\Sigma}_{z}}%
}\right)  \mathbf{\rho}}\nonumber\\
&  \exp\left[  -\frac{n}{4}\mathrm{STr}\left(  (1+\mathrm{e}^{-z/2}%
)\mathbf{Q+}(1-\mathrm{e}^{-z/2})\mathbf{\mathbf{\Sigma}_{z}Q\mathbf{\Sigma
}_{z}}\right)  ^{2}\right]  ,\nonumber
\end{align}
where we have employed the fact that in this representation $\mathbf{Q}%
\rightarrow\frac{1}{2}(\mathbf{\mathbf{\Sigma}_{z}Q}+\mathbf{Q\mathbf{\Sigma
}_{z}})+\frac{1}{2}\mathrm{e}^{-z/2}(\mathbf{Q}-\mathbf{\mathbf{\Sigma}%
_{z}Q\mathbf{\Sigma}_{z}})$ and $\mathbf{Q}^{\prime}\rightarrow-\sqrt
{\mathbf{\mathbf{\Sigma}_{z}}}\mathbf{Q}\sqrt{\mathbf{\mathbf{\Sigma}_{z}}}$.

The main features of the tunneling conductance can be found in the framework
of the simplest version of this technique, which ignores the coordinate
dependence $Q_{i,j}(r)=Q_{i,j}$ of the matrix elements Eq.(\ref{Q}).\cite{IZ}
It is applicable when the perturbation of the mean field regime by the
condensate density fluctuations is strong enough, in complete analogy with the
situation known in Quantum Mechanics, where a strong time-dependent
perturbation induce quantum transitions between slightly detuned levels as if
it were resonant. In such a situation, only the integral of the time
dependence enters the result, while the particular form of the time dependence
plays no role and can be taken uniform. Here, by the analogy, we have to
compere the interactions resulting in two last terms in the exponent in the
second line of Eq.(\ref{ProbDQ}). Since we consider tunneling at a long
distance $R,$ the typical difference of the momenta of the advanced and
retarded Green's functions is small, and according to Eq.(\ref{Interaction})
the typical variation of the interaction associated with the term
$\mathrm{STr}n\mathbf{\mathbf{\Sigma}_{z}V}$ can be estimated as
$\mathcal{N}\mathbf{\partial}_{p}V(p)R^{-1}\mathbf{\sim}\mathcal{N}\Lambda
R^{-1}$. The fluctuation-induced coupling associated with the term
$\mathrm{STr}n\mathbf{VQ}^{\prime}$ can be estimated from the condition
$\frac{n}{4}\mathrm{STr}\mathbf{Q}^{2}\sim\frac{n\mathcal{V}}{4}%
\mathrm{Str}\widehat{Q}^{2}\sim1$, that is $\widehat{Q}\sim1/\sqrt
{\mathcal{N}}$, and hence this coupling is of the order of $\Lambda
\sqrt{\mathcal{N}}$. Therefore, the fluctuations can be considered as uniform
for $\mathcal{N}R^{-2}\sim nR\ll1$, which we assume to be the case for the
system under consideration.

The assumption of spacial uniform matrix elements allows one to employ the
momentum representation where two momenta $p$ and $q$ corresponding to the
advanced and retarded Green's functions, respectively. \ This simplifies
considerably the integral over $\mathcal{D}\mathbf{Q}(r)$ in Eq.(\ref{ProbDQ}%
), \ which now becomes not a functional, but just an $8$-fold integral over
the regular and an $8$-fold integral over the Grassmann variables%
\begin{align}
&  P(\zeta,\eta,z)=n^{2}\int\frac{\mathrm{d}p\mathrm{d}q}{\left(  2\pi\right)
^{2d}}~\mathrm{e}^{\mathrm{i}R\left(  p-q\right)  }\int\mathcal{D}%
\overline{\phi}\mathcal{D}\phi\nonumber\\
&  \int\mathrm{d}\widehat{Q}\mathrm{e}^{-\frac{\mathcal{N}}{4}\mathrm{STr}%
\left(  (1+\mathrm{e}^{-z/2})\widehat{Q}\mathbf{+}(1-\mathrm{e}^{-z/2}%
)\widehat{\Sigma}_{z}\widehat{Q}\widehat{\Sigma}_{z}\right)  ^{2}%
}\label{ProbMomentum}\\
&  \overline{\lambda}_{f}(p)\lambda_{f}(p)\mu_{f}(q)\overline{\mu}%
_{f}(q)\nonumber\\
&  \mathrm{e}~^{\int\frac{\mathcal{V}\mathrm{d}p^{\prime}}{\left(
2\pi\right)  ^{d}}\phi(p^{\prime})\left(  \mathrm{i}\mathbf{\zeta+}%
\mathrm{i}\eta\widehat{\Sigma}\mathbf{_{z}+}\mathrm{i}n\widehat{\Sigma
}\mathbf{_{z}}V(p^{\prime})\mathbf{-}\mathrm{i}nV(p^{\prime})\widehat
{Q}^{\prime}\right)  \overline{\phi}(p^{\prime})}.\nonumber
\end{align}
The pre-exponential factors can be written as derivatives%
\begin{align}
\overline{\lambda}_{f}(p)\lambda_{f}(p)  &  \rightarrow\frac{\partial
_{J_{a}\rightarrow0}}{\mathrm{i}~nV(p)}\mathrm{e}^{\mathrm{i}~\int
\frac{\mathcal{V}\mathrm{d}p^{\prime}}{\left(  2\pi\right)  ^{d}}nV(p^{\prime
})~\phi(p^{\prime})\widehat{J}_{a}\overline{\phi}(p^{\prime})};\nonumber\\
\widehat{J}_{a}  &  =J_{a}\widehat{j}_{a}\frac{\left(  2\pi\right)  ^{d}%
}{\mathcal{V}}\delta\left(  p-p^{\prime}\right) \label{deltaa}\\
\widehat{j}_{a}  &  =\left(
\begin{array}
[c]{cccc}%
0 & 0 & 0 & 0\\
0 & 0 & 0 & 0\\
0 & 0 & 1 & 0\\
0 & 0 & 0 & 0
\end{array}
\right)
\end{align}
and%
\begin{align}
\mu_{f}(q)\overline{\mu}_{f}(q)  &  \rightarrow\frac{\partial_{J_{b}%
\rightarrow0}}{\mathrm{i}~nV(q)}\mathrm{e}^{\mathrm{i}~\int\frac
{\mathcal{V}\mathrm{d}p^{\prime}}{\left(  2\pi\right)  ^{d}}nV(p^{\prime
})~\phi(p^{\prime})\widehat{J}_{a}\overline{\phi}(p^{\prime})};\nonumber\\
\widehat{J}_{b}  &  =J_{b}\widehat{j}_{b}\frac{\left(  2\pi\right)  ^{d}%
}{\mathcal{V}}\delta\left(  q-p^{\prime}\right)  ,\label{deltar}\\
\widehat{j}_{b}  &  =\left(
\begin{array}
[c]{cccc}%
0 & 0 & 0 & 0\\
0 & 0 & 0 & 0\\
0 & 0 & 0 & 0\\
0 & 0 & 0 & 1
\end{array}
\right)
\end{align}
which allows one to write Eq.(\ref{ProbMomentum}) in the form%
\begin{align}
&  P(\zeta,\eta,z)=\partial_{J_{a}J_{b}\rightarrow0}^{2}\int\frac
{\mathrm{d}p\mathrm{d}q}{\left(  2\pi\right)  ^{2d}}\mathrm{e}^{\mathrm{i}%
R\left(  p-q\right)  }\int\frac{\mathcal{D}\overline{\phi}\mathcal{D}\phi
}{V(p)V(q)}\label{ChangeOrder}\\
&  \int\mathrm{d}\widehat{Q}\mathrm{e}^{-\frac{\mathcal{N}}{4}\mathrm{STr}%
\left(  (1+\mathrm{e}^{-z/2})\widehat{Q}\mathbf{+}(1-\mathrm{e}^{-z/2}%
)\widehat{\Sigma}_{z}\widehat{Q}\widehat{\Sigma}_{z}\right)  ^{2}}\nonumber\\
&  \mathrm{e}~^{\int\frac{\mathrm{i}\mathcal{V}\mathrm{d}p^{\prime}}{\left(
2\pi\right)  ^{d}}\phi(p)\left(  \mathbf{\zeta+\mathrm{i}}\eta\widehat{\Sigma
}\mathbf{_{z}+}n\widehat{\Sigma}\mathbf{_{z}}V(p)\mathbf{+}nV(p)\left(
\widehat{Q}^{\prime}-\widehat{J}_{a}+\widehat{J}_{b}\right)  \right)
\overline{\phi}(p)}.\nonumber
\end{align}
One can substitute now explicit expression Eq.(\ref{Interaction}) for the
interaction potential, (we employ the scaled units $\ \eta\rightarrow\eta
n\Lambda,$ $\ $\ ${\normalsize \zeta\rightarrow\zeta n\Lambda,}t\rightarrow
t/n\Lambda$\ for $t$, \ $\eta$ $\ $and \ $\mathbf{\zeta}$, respectively,
\ that have been introduced in Eq.(\ref{1.7})), and consider domain $\eta
\sim-1$ by making the transformation $\eta\rightarrow\eta-1$, along with trhe
assumption $\eta\ll1$ implied by the physical requirement $p,$ $q<<\kappa
$.\ This yields%
\begin{align*}
&  P(\zeta,\eta,z)=\partial_{J_{a}J_{b}\rightarrow0}^{2}\int\frac
{\mathrm{d}p\mathrm{d}q}{\left(  2\pi\right)  ^{2d}}\mathrm{e}^{\mathrm{i}%
R\left(  p-q\right)  }\int\mathcal{D}\overline{\phi}\mathcal{D}\phi\\
&  \int\mathrm{d}\widehat{Q}\mathrm{e}^{-\frac{\mathcal{N}}{4}\mathrm{STr}%
\left(  (1+\mathrm{e}^{-z/2})\widehat{Q}\mathbf{+}(1-\mathrm{e}^{-z/2}%
)\widehat{\Sigma}_{z}\widehat{Q}\widehat{\Sigma}_{z}\right)  ^{2}}\\
&  \mathrm{e}~^{\int\frac{\mathcal{V}\mathrm{d}p}{\left(  2\pi\right)  ^{d}%
}\phi(p^{\prime})\left(  \mathrm{i}\mathbf{\zeta+}\mathrm{i}\left(
\eta-p^{\prime2}\right)  \widehat{\Sigma}\mathbf{_{z}-}\mathrm{i}~\left(
\widehat{Q}^{\prime}-\widehat{J}_{a}+\widehat{J}_{b}\right)  \right)
\overline{\phi}(p^{\prime})},
\end{align*}
where $\zeta$ and $\eta$ were neglected when compared to unity. After having
shifted the matrix $\widehat{Q}$ \ by $\mathbf{\zeta}\widehat{\Sigma
}\mathbf{\mathbf{_{z}}}$, one obtains a form of this expression%
\begin{align}
&  P(\zeta,\eta,z)=\partial_{J_{a}J_{b}\rightarrow0}^{2}\int\frac
{\mathrm{d}p\mathrm{d}q}{\left(  2\pi\right)  ^{2d}}\mathrm{e}^{\mathrm{i}%
R\left(  p-q\right)  }\int\mathcal{D}\overline{\phi}\mathcal{D}\phi
\label{ChOr}\\
&  \int\mathrm{d}\widehat{Q}\mathrm{e}^{-\frac{\mathcal{N}}{2}\left(
4\zeta\mathrm{Str}\widehat{\Sigma}\mathbf{\mathbf{_{z}}}\widehat
{Q}+(1+\mathrm{e}^{-z})\mathrm{Str}\widehat{Q}^{2}+(1-\mathrm{e}%
^{-z})\mathrm{Str}\left(  \widehat{Q}\widehat{\Sigma}\mathbf{_{z}}\right)
^{2}\right)  }\nonumber\\
&  \mathrm{e}~^{\int\frac{\mathcal{V}\mathrm{d}p^{\prime}}{\left(
2\pi\right)  ^{d}}\phi(p^{\prime})\left(  \mathrm{i}\left(  \eta-p^{\prime
2}\right)  \widehat{\Sigma}\mathbf{_{z}-}\mathrm{i}\left(  \widehat{Q}%
^{\prime}-\widehat{J}_{a}+\widehat{J}_{b}\right)  \right)  \overline{\phi
}(p^{\prime})},\nonumber
\end{align}
convenient for further calculations.

We did not yet specified the precise form of the matrix $\widehat{Q}$ that
allows one to change order of the integration in Eq.(\ref{ChOr}) and evaluate
the integral over $\mathcal{D}\overline{\phi}\mathcal{D}\phi$ first. The
standard approach suggests to express this matrix in terms of the generators
of the superalgebra $osp(1,1|2)$\ and make use of the parametrization
\begin{equation}
\widehat{Q}=\widehat{T}\widehat{P}\widehat{T}^{-1}, \label{Qpar}%
\end{equation}
where $\widehat{P}$ is a supermatrix
\begin{equation}
\widehat{P}=\left(
\begin{array}
[c]{cccc}%
q_{1} & 0 & \overline{\kappa}_{1} & 0\\
0 & q_{2} & 0 & \overline{\kappa}_{2}\\
\kappa_{1} & 0 & \mathrm{i}\widetilde{q}_{1} & 0\\
0 & \kappa_{2} & 0 & \mathrm{i}\widetilde{q}_{2}%
\end{array}
\right)  \label{P}%
\end{equation}
and $\widehat{T}$ is a pseudounitary rotation%
\[
\widehat{T}=\widehat{U}\widehat{M}%
\]
with%
\begin{align*}
\widehat{M}=  &  \left(
\begin{array}
[c]{cccc}%
\cosh\Theta & \mathrm{e}^{\mathrm{i}\Phi}\sinh\Theta & 0 & 0\\
\mathrm{e}^{-\mathrm{i}\Phi}\sinh\Theta & \cosh\Theta & 0 & 0\\
0 & 0 & \cos\theta & \mathrm{ie}^{\mathrm{i}\phi}\sin\theta\\
0 & 0 & \mathrm{ie}^{-\mathrm{i}\phi}\sin\theta & \cos\theta
\end{array}
\right) \\
&  \widehat{U}=\left(
\begin{array}
[c]{cccc}%
1-\frac{\overline{\alpha}\alpha}{2} & 0 & -\overline{\alpha} & 0\\
0 & 1+\frac{\overline{\beta}\beta}{2} & 0 & -\mathrm{i}\overline{\beta}\\
\alpha & 0 & 1+\frac{\overline{\alpha}\alpha}{2} & 0\\
0 & \mathrm{i}\beta & 0 & 1-\frac{\overline{\beta}\beta}{2}%
\end{array}
\right)  .
\end{align*}
In this representation the integration measure (Haar measure for the chosen
parametrization) reads%
\begin{align}
\mathrm{d}\widehat{Q}  &  =\mathrm{d}\widehat{P}\mathrm{d}\widehat
{T}\label{Haar}\\
\mathrm{d}\widehat{T}  &  =\mathrm{d}\cosh2\Theta~\mathrm{d}\cos2\theta
\frac{\mathrm{d}\Phi\mathrm{d}\varphi}{4\pi^{2}}\frac{\mathrm{d}%
\alpha\mathrm{d}\beta\mathrm{d}\overline{\alpha}\mathrm{d}\overline{\beta}%
}{\left(  \cosh2\Theta~-\cos2\theta\right)  ^{2}}\nonumber\\
\mathrm{d}\widehat{P}  &  =\mathrm{d}q_{1}\mathrm{d}q_{2}\mathrm{d}%
\widetilde{q}_{1}\mathrm{d}\widetilde{q}_{2}\mathrm{d}\kappa_{1}%
\mathrm{d}\kappa_{2}\mathrm{d}\overline{\kappa}_{1}\mathrm{d}\overline{\kappa
}_{2}.\nonumber
\end{align}

We apply this parametrization to the generating function in the form
Eq.(\ref{ProbDQ1}) and perform the transformations Eqs.(\ref{ProbMomentum}%
-\ref{ChangeOrder}). Evaluation of the integral over $\mathcal{D}%
\overline{\phi}\mathcal{D}\phi$ yields%
\begin{align*}
&  P(\zeta,\eta,z)=\partial_{J_{a},J_{b}\rightarrow0}^{2}\int\frac
{\mathrm{d}p\mathrm{d}q}{\left(  2\pi\right)  ^{2d}}~\mathrm{e}^{\mathrm{i}%
R\left(  p-q\right)  }\\
&  \int\mathrm{d}\widehat{Q}\mathrm{e}^{-\mathcal{N}\left(  \frac
{1+\mathrm{e}^{-z}}{2}\mathrm{Str}\widehat{Q}^{2}+\frac{1-\mathrm{e}^{-z}}%
{2}\mathrm{Str}\left(  \widehat{Q}\widehat{\Sigma}_{z}\right)  ^{2}%
-\zeta\mathrm{Str}\widehat{\Sigma}_{z}\widehat{Q}\right)  }\\
&  \mathrm{e}^{-\int\frac{\mathcal{N}\mathrm{d}p^{\prime}}{2\left(
2\pi\right)  ^{d}}\mathrm{Str\ln}{\Huge [}\eta-p^{\prime2}\mathbf{-}%
\widehat{Q}^{\prime}+\left(  \widehat{J}_{a}-\widehat{J}_{b}\right)
\widehat{\Sigma}_{z}{\Huge ]}}.
\end{align*}
By substituting $\widehat{Q}^{\prime}=-\sqrt{\widehat{\Sigma}_{z}}\widehat
{Q}\sqrt{\widehat{\Sigma}_{z}}$ and by moving $\sqrt{\widehat{\Sigma}_{z}}$
under the supertrace operators one finds%
\begin{align}
&  P(\zeta,\eta,z)=\partial_{J_{a},J_{b}\rightarrow0}^{2}\int\frac
{\mathrm{d}p\mathrm{d}q}{\left(  2\pi\right)  ^{2d}}~\mathrm{e}^{\mathrm{i}%
R\left(  p-q\right)  }\nonumber\\
&  \int\mathrm{d}\widehat{Q}\mathrm{e}^{-\mathcal{N}\left(  \frac
{1+\mathrm{e}^{-z}}{2}\mathrm{Str}\widehat{Q}^{2}+\frac{1-\mathrm{e}^{-z}}%
{2}\mathrm{Str}\left(  \widehat{Q}\widehat{\Sigma}_{z}\right)  ^{2}%
-\zeta\mathrm{Str}\widehat{\Sigma}_{z}\widehat{Q}\right)  }\label{ProbScal}\\
&  \mathrm{e}^{-\int\frac{\mathcal{V}\mathrm{d}p^{\prime}}{2\left(
2\pi\right)  ^{d}}\mathrm{Str\ln}{\Huge [}\eta-p^{\prime2}\mathbf{-}%
\widehat{Q}+\left(  \widehat{J}_{a}-\widehat{J}_{b}\right)  {\Huge ]}%
},\nonumber
\end{align}
which after taking the derivatives%
\begin{align*}
&  P(\zeta,\eta,z)=\frac{n^{2}}{4}~\int\mathrm{d}\widehat{Q}\mathrm{e}%
^{-\int\frac{\mathcal{V}\mathrm{d}p^{\prime}}{2\left(  2\pi\right)  ^{d}%
}\mathrm{Str\ln}{\Huge [}\eta-p^{\prime2}\mathbf{-}\widehat{Q}{\Huge ]}}\\
&  \mathrm{e}^{\mathcal{N}\zeta\mathrm{Str}\widehat{\Sigma}_{z}\widehat
{Q}-\mathcal{N}\frac{1+\mathrm{e}^{-z}}{2}\mathrm{Str}\widehat{Q}%
^{2}-\mathcal{N}\frac{1-\mathrm{e}^{-z}}{2}\mathrm{Str}\left(  \widehat
{Q}\widehat{\Sigma}_{z}\right)  ^{2}}\\
&  \int\frac{\mathrm{d}p\mathrm{d}q}{\left(  2\pi\right)  ^{2d}}%
\mathrm{e}^{\mathrm{i}R\left(  p-q\right)  }\mathrm{Str}\widehat{j}_{a}%
\frac{1}{\eta-p^{2}\mathbf{-}\widehat{Q}}\mathrm{Str}\widehat{j}_{b}\frac
{1}{\eta-q^{2}\mathbf{-}\widehat{Q}}%
\end{align*}
allows one to explicitly perform the integration over $\mathrm{d}p\mathrm{d}q$
and results in
\begin{align}
&  P(\zeta,\eta,z)=~\int\mathrm{d}\widehat{Q}\mathrm{e}^{-\int\frac
{\mathcal{V}\mathrm{d}p^{\prime}}{2\left(  2\pi\right)  ^{d}}\mathrm{Str\ln
}{\Huge [}\eta-p^{\prime2}\mathbf{-}\widehat{Q}{\Huge ]}}\label{GenEx}\\
&  \mathrm{e}^{\mathcal{N}\zeta\mathrm{Str}\widehat{\Sigma}_{z}\widehat
{Q}-\mathcal{N}\frac{1+\mathrm{e}^{-z}}{2}\mathrm{Str}\widehat{Q}%
^{2}-\mathcal{N}\frac{1-\mathrm{e}^{-z}}{2}\mathrm{Str}\left(  \widehat
{Q}\widehat{\Sigma}_{z}\right)  ^{2}}\nonumber\\
&  \mathrm{Str}\widehat{j}_{a}G\left(  \widehat{Q}\right)  ~~\mathrm{Str}%
\widehat{j}_{b}G\left(  \widehat{Q}\right)  ~,\nonumber
\end{align}
where by the analogy to Eq.(\ref{1.10}) the Green's functions (depending on
the matrix $\widehat{Q}$) read
\[%
\begin{array}
[c]{c}%
G_{A,R}\left(  \widehat{Q}\right)  =\frac{n\mathrm{e}^{\pm\mathrm{i}%
R\sqrt{\eta\mathbf{-}\widehat{Q}}}}{2\sqrt{\eta\mathbf{-}\widehat{Q}}%
}~~~~~~~~~~~~~~~\mathrm{~for~1D}\\
G_{A,R}\left(  \widehat{Q}\right)  =\frac{nK_{0}(\pm\mathrm{i}R\sqrt
{\eta\mathbf{-}\widehat{Q}})}{2\pi}~~~~~~~\mathrm{for~2D}\\
G_{A,R}\left(  \widehat{Q}\right)  =\frac{n\mathrm{e}^{\pm\mathrm{i}%
R\sqrt{\eta\mathbf{-}\widehat{Q}}}}{4\pi R}~~~~~~~~~~~~~~~~\mathrm{for~3D,}%
\end{array}
\]
and the choice of signs has to be done such that to assure the convergency.

By moving $\widehat{T}$ under the supertraces and superdeterminants, one can
obtain Eq.(\ref{GenEx}) in the form%
\begin{align}
&  P(\zeta,\eta,z)=~\int\mathrm{d}\widehat{P}\mathrm{d}\widehat{T}%
\mathrm{e}^{-\int\frac{\mathcal{V}\mathrm{d}p^{\prime}}{2\left(  2\pi\right)
^{d}}\mathrm{Str\ln}{\Huge [}\eta-p^{\prime2}\mathbf{-}\widehat{P}{\Huge ]}%
}\nonumber\\
&  \mathrm{e}^{-\mathcal{N}\frac{\zeta}{2}\mathrm{Str}\widehat{\Sigma}%
_{z}^{\prime}\widehat{P}-\mathcal{N}\frac{1+\mathrm{e}^{-z}}{2}\mathrm{Str}%
\widehat{P}^{2}-\mathcal{N}\frac{1-\mathrm{e}^{-z}}{2}\mathrm{Str}\left(
\widehat{P}\widehat{\Sigma}_{z}^{\prime}\right)  ^{2}}\nonumber\\
&  \mathrm{Str}\widehat{j}_{a}G\left(  \widehat{P}\right)  ~~\mathrm{Str}%
\widehat{j}_{b}G\left(  \widehat{P}\right)  \label{GenP}%
\end{align}
convenient for integration over the matrix\textrm{ }$\widehat{P}$ . Here the
integration over $\mathrm{d}\widehat{P}\mathrm{d}\widehat{T}$ has to be done
according to the measure Eq.(\ref{Haar}), the matrices with primes read
\begin{widetext}%
\begin{align}
\mathbf{\widehat{\Sigma}}^{\prime}\mathbf{\mathbf{_{z}}}  &  =\left(
\begin{array}
[c]{cccc}%
\cosh2\Theta & -\mathrm{e}^{2\mathrm{i}\Phi}\sinh2\Theta & 0 & 0\\
\mathrm{e}^{-2\mathrm{i}\Phi}\sinh2\Theta & -\cosh2\Theta & 0 & 0\\
0 & 0 & \cos2\theta & -\mathrm{ie}^{-2\mathrm{i}\phi}\sin2\theta\\
0 & 0 & \mathrm{ie}^{2\mathrm{i}\phi}\sin2\theta & -\cos2\theta
\end{array}
\right)  ,\label{Sigma}\\
\widehat{j}_{a}^{\prime}  &  =\left(
\begin{array}
[c]{cccc}%
\overline{\alpha}\alpha\cosh^{2}\Theta & -\frac{\overline{\alpha}\alpha}%
{2}\mathrm{e}^{2\mathrm{i}\Phi}\sinh2\Theta & -\overline{\alpha}\cosh
\Theta\cos\theta & \overline{\alpha}\mathrm{ie}^{-2\mathrm{i}\phi}\cosh
\Theta\sin\theta\\
\frac{\overline{\alpha}\alpha}{2}\mathrm{e}^{-2\mathrm{i}\Phi}\sinh2\Theta &
-\overline{\alpha}\alpha\sinh^{2}\Theta & -\overline{\alpha}\sinh
\Theta\mathrm{e}^{-2\mathrm{i}\Phi}\cos\theta & \overline{\alpha}%
\mathrm{ie}^{-2\mathrm{i}\left(  \Phi+\phi\right)  }\sinh\Theta\sin\theta\\
-\alpha\cosh\Theta\cos\theta & \alpha\sinh\Theta\mathrm{e}^{2\mathrm{i}\Phi
}\cos\theta & \left(  1+\overline{\alpha}\alpha\right)  \cos^{2}\theta &
-\mathrm{ie}^{-2\mathrm{i}\phi}\frac{1+\overline{\alpha}\alpha}{2}\sin
2\theta\\
-\alpha\mathrm{ie}^{2\mathrm{i}\phi}\cosh\Theta\sin\theta & \alpha
\mathrm{ie}^{2\mathrm{i}\left(  \Phi+\phi\right)  }\sinh\Theta\sin\theta &
\mathrm{ie}^{2\mathrm{i}\phi}\frac{1+\overline{\alpha}\alpha}{2}\sin2\theta &
\left(  1+\overline{\alpha}\alpha\right)  \sin^{2}\theta
\end{array}
\right) \label{jpa}\\
\widehat{j}_{b}^{\prime}  &  =\left(
\begin{array}
[c]{cccc}%
\overline{\beta}\beta\cosh^{2}\Theta & -\frac{\overline{\beta}\beta}%
{2}\mathrm{e}^{2\mathrm{i}\Phi}\sinh2\Theta & -\overline{\beta}\mathrm{e}%
^{2\mathrm{i}\left(  \Phi+\phi\right)  }\sinh\Theta\sin\theta & -\overline
{\beta}\mathrm{ie}^{2\mathrm{i}\Phi}\sinh\Theta\cos\theta\\
\frac{\overline{\beta}\beta}{2}\mathrm{e}^{-2\mathrm{i}\Phi}\sinh2\Theta &
-\overline{\beta}\beta\sinh^{2}\Theta & -\overline{\beta}\mathrm{e}%
^{2\mathrm{i}\phi}\cosh\Theta\sin\theta & -\overline{\beta}\mathrm{i}%
\cosh\Theta\cos\theta\\
-\beta\mathrm{e}^{-2\mathrm{i}\left(  \Phi+\phi\right)  }\sinh\Theta\sin\theta
& \beta\mathrm{e}^{-2\mathrm{i}\phi}\cosh\Theta\sin\theta & \left(
1-\overline{\beta}\beta\right)  \sin^{2}\theta & \mathrm{ie}^{-2\mathrm{i}%
\phi}\frac{1-\overline{\beta}\beta}{2}\sin2\theta\\
\beta\mathrm{ie}^{-2\mathrm{i}\Phi}\sinh\Theta\cos\theta & -\beta
\mathrm{i}\cosh\Theta\cos\theta & -\mathrm{ie}^{2\mathrm{i}\phi}%
\frac{1-\overline{\beta}\beta}{2}\sin2\theta & \left(  1-\overline{\beta}%
\beta\right)  \cos^{2}\theta
\end{array}
\right)  , \label{jpb}%
\end{align}
and the\ terms in the exponent have the explicit form%
\begin{align}
\mathrm{Str}\widehat{\Sigma}^{\prime}\mathbf{\mathbf{_{z}}}\widehat{P}  &
=\left(  q_{1}-q_{2}\right)  \cosh2\Theta-\mathrm{i}\left(  \widetilde{q}%
_{1}-\widetilde{q}_{2}\right)  \cos2\theta\label{expp}\\
\mathrm{Str}\widehat{P}^{2}  &  =q_{1}^{2}+q_{2}^{2}+\widetilde{q}_{1}%
^{2}+\widetilde{q}_{2}^{2}+2\overline{\kappa}_{1}\kappa_{1}+2\overline{\kappa
}_{2}\kappa_{2}\nonumber\\
\mathrm{Str}\left(  \widehat{P}\widehat{\Sigma}_{z}^{\prime}\right)  ^{2}  &
=\left(  q_{1}^{2}+q_{2}^{2}\right)  \cosh^{2}2\Theta-2q_{1}q_{2}\sinh
^{2}2\Theta+\left(  \widetilde{q}_{1}^{2}+\widetilde{q}_{2}^{2}\right)
\cos^{2}2\theta+2\widetilde{q}_{1}\widetilde{q}_{2}\sin^{2}2\theta+\nonumber\\
&  2\left(  \overline{\kappa}_{1}\kappa_{1}+\overline{\kappa}_{2}\kappa
_{2}\right)  \cosh2\Theta\cos2\theta-2\mathrm{i}\left(  \mathrm{e}%
^{2\mathrm{i}\left(  \Phi+\phi\right)  }\overline{\kappa}_{2}\kappa
_{1}+\mathrm{e}^{-2\mathrm{i}\left(  \Phi+\phi\right)  }\overline{\kappa}%
_{1}\kappa_{2}\right)  \sinh2\Theta\sin2\theta.\nonumber
\end{align}%
\end{widetext}%

The integral over $\mathrm{d}\widehat{P}$ can be evaluated by the saddle point
method. The main contribution, generally speaking, comes from a saddle point
where the matrix elements $q_{1},q_{2},\widetilde{q}_{1},\widetilde{q}_{2}$ of
$\widehat{P}$ are shifted to the complex plane, whence the matrix $\widehat
{P}$ has a diagonal form $\widehat{P_{s}}$ no longer corresponding to
Eq.(\ref{P}) and independent on Grassmann variables. The latter, as well as
the deviations of the regular variables are considered as perturbations. Note
that the pre-expnential factor does not change as the result of integration
over these perturbations, since the Grassmann integration around
$\widehat{P_{s}}$ compensate for the pre-factor resulting from the integration
over the regular deviations. Also note that the among different possibilities,
the saddle point has to be chosen in such a way that the integral over
$\mathrm{d}\widehat{T}$ converges. Moreover, the choice $q_{1}=-q_{2}%
=\mathrm{i}\widetilde{q}_{1}=-\mathrm{i}\widetilde{q}_{2}$ $=\mathrm{i}q_{s}$
at the saddle point allows one to get rid of the large terms $\mathcal{N}%
\mathrm{Str}\widehat{P_{s}}^{2}$ and $\mathrm{STr\ln}{\Huge [}\eta-p^{\prime
2}\mathbf{-}\widehat{P_{s}}{\Huge ]}$, and Eq.(\ref{GenP}) takes the form%
\begin{align}
&  P(\zeta,\eta,z)=n^{2}~\int\mathrm{d}\widehat{T}\mathrm{e}^{-2\mathcal{N}%
\zeta\mathrm{i}q_{s}\left(  \cosh2\Theta-\cos2\theta\right)  }\nonumber\\
&  \mathrm{e}^{-2\mathcal{N}\left(  1-\mathrm{e}^{-z}\right)  q_{s}^{2}\left(
\cosh^{2}2\Theta-\cos^{2}2\theta\right)  }\nonumber\\
&  \mathrm{Str}\widehat{j}_{a}G_{A,R}\left(  \widehat{P_{s}}\right)
~~\mathrm{Str}\widehat{j}_{b}G_{A,R}\left(  \widehat{P_{s}}\right)
\label{GenSad}%
\end{align}
where the choice between the advanced and retarded Green's functions depends
on the position of the matrix element $q_{i}=\pm\mathrm{i}q_{s}$ on the
complex plane.

Since $\mathrm{Str}\widehat{j}_{a}^{\prime}\widehat{G}=\alpha\overline{\alpha
}(\cosh2\Theta-\cos2\theta)\frac{G_{R}-G_{A}}{2}-\sin^{2}\theta(G_{R}-G_{A})$
\ and $\mathrm{Str}\widehat{j}_{b}^{\prime}\widehat{G}=\beta\overline{\beta
}(\cosh2\Theta-\cos2\theta)\frac{G_{R}-G_{A}}{2}+\sin^{2}\theta(G_{R}-G_{A}%
)$\ by integration over $\mathrm{d}\Phi\mathrm{d}\phi\mathrm{d}\alpha
\mathrm{d}\beta\mathrm{d}\overline{\alpha}\mathrm{d}\overline{\beta}$ , one
can explicitly find the pre-exponential factors in Eq.(\ref{GenSad})%
\[
\mathrm{Str}\widehat{j}_{a}^{\prime}\widehat{G}~~\mathrm{Str}\widehat{j}%
_{b}^{\prime}\widehat{G}~=-\frac{1}{\pi^{2}}(\cosh2\Theta-\cos2\theta
)^{2}(G_{R}-G_{A})^{2}%
\]
with the Green's functions%
\begin{equation}%
\begin{array}
[c]{c}%
G_{A}\rightarrow\frac{n\mathrm{e}^{-\mathrm{i}R\sqrt{\eta\mathbf{+}%
\mathrm{i}q_{s}}}}{2\sqrt{\eta\mathbf{+}\mathrm{i}q_{s}}}~~G_{R}%
\rightarrow\frac{n\mathrm{e}^{\mathrm{i}R\sqrt{\eta\mathbf{-}q_{s}}}}%
{2\sqrt{\eta\mathbf{-}\mathrm{i}q_{s}}}~~~~~~~~~~~~~~\mathrm{~for~1D}\\
G_{A}\rightarrow\frac{nK_{0}(-\mathrm{i}R\sqrt{\eta\mathbf{+}\mathrm{i}q_{s}%
})}{2\pi}~~~G_{R}\rightarrow\frac{nK_{0}(\mathrm{i}R\sqrt{\eta\mathbf{-}%
\mathrm{i}q_{s}})}{2\pi}~~\mathrm{for~2D}\\
G_{A}\rightarrow\frac{n\mathrm{e}^{-\mathrm{i}R\sqrt{\eta\mathbf{+}%
\mathrm{i}q_{s}}}}{4\pi R}~~~G_{R}\rightarrow\frac{n\mathrm{e}^{\mathrm{i}%
R\sqrt{\eta\mathbf{-}\mathrm{i}q_{s}}}}{4\pi R}~~~~~~~~~~~~~\mathrm{for~3D.}%
\end{array}
\label{GRan}%
\end{equation}
The replacement%
\[
\cosh2\Theta\rightarrow Z,\cos2\theta\rightarrow w
\]
followed by simple algebra makes the explicit expressions Eq.(\ref{GenSad})
for the generating function more compact%
\begin{align}
&  P(\zeta,\eta,z)=~\int\limits_{1}^{\infty}\mathrm{d}Z\int\limits_{-1}%
^{1}~\mathrm{d}w\mathrm{e}^{-\mathcal{N}\frac{1-\mathrm{e}^{-z}}{2}q_{s}%
^{2}\left(  Z^{2}-w^{2}\right)  }\nonumber\\
&  \mathrm{e}^{-2\mathcal{N}\zeta\mathrm{i}q_{s}\left(  Z-w\right)  }%
(G_{A}-G_{R})^{2}. \label{GenZ}%
\end{align}
At the limit $z\rightarrow0$ the position of the saddle point is independent
from $Z$ and $w$, which immediately yields the tunneling probability%
\begin{equation}
P(\zeta,\eta,0)=\frac{(G_{A}-G_{R})^{2}}{2}\int\limits_{1}^{\infty}%
\mathrm{d}Z\int\limits_{-1}^{1}~\mathrm{d}w\mathrm{e}^{-\mathcal{N}%
\zeta\mathrm{i}q_{s}\left(  Z-w\right)  }, \label{ProbZ}%
\end{equation}
and the number of the atoms%
\begin{align}
&  P_{z}^{\prime}(\zeta,\eta,0)=~\frac{q_{s}^{2}\mathcal{N}}{2}(G_{A}%
-G_{R})^{2}\label{dNZ}\\
&  \int\limits_{1}^{\infty}\mathrm{d}Z\int\limits_{-1}^{1}~\mathrm{d}%
w\mathrm{e}^{-\mathcal{N}\zeta\mathrm{i}q_{s}\left(  Z-w\right)  }\left(
Z^{2}-w^{2}\right) \nonumber
\end{align}
that have been kicked out from the condensate as the result of the tunneling.

\begin{figure}
[ptb]
\begin{center}
\includegraphics[
height=1.2341in,
width=2.7873in
]%
{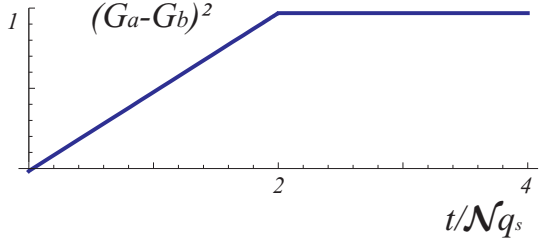}%
\caption{Time dependence(not to scale), corresponding to the frequency
dependence of the tunneling probability for $\gamma=0$.}%
\label{FourierCoff}%
\end{center}
\end{figure}

Calculation of the integrals Eq.(\ref{ProbZ}) for the probability results in%
\begin{equation}
P(\zeta,\eta,0)=~-\left(  G_{A}-G_{R}\right)  {}^{2}\frac{\sin^{2}\left(
\mathcal{N}\zeta q_{s}\right)  }{2\mathcal{N}^{2}\zeta^{2}q_{s}^{2}},
\label{TunProbS}%
\end{equation}
It is expedient to show in Fig.\ref{FourierCoff} the Fourier transforms over
the frequency $\zeta$ of the coefficients in front of the $\left(  G_{A}%
-G_{R}\right)  {}^{2}$ in Eq.(\ref{TunProbS}).%
The electron tunnelling probabilitythrough the condensate linearly grows as
function of time and saturates at the Heisenberg time, which is a natural
consequence of the employed $0D$ model of fluctuations (coordinate independent
matrix $\mathbf{Q}$). However, the typical saturation time $t\sim
\mathcal{N}q_{s}$ depends on the energy $\eta$ via the energy dependence of
the position $q_{s}(\eta)$ of the saddle point position in the complex plane,
which we will find later on. Note that, a rapidly oscillating part of the
coordinate dependence, which is present due to the interference beats between
$G_{A}$ and $G_{R}$ for the leads size small compared to $1/\sqrt{\eta}$, must
disapear for large liads, and the combination $\left(  G_{A}-G_{R}\right)
{}^{2}$ has to be replaced by $2G_{A}G_{R}$.

Let us now calculate the number of atoms kicked out of the condensate. At this
stage we have to take into account the fact that the frequency $\zeta
=\omega-\mathrm{i}\gamma$ has an imaginary part $-\mathrm{i}\gamma$
originating from the fact that the electron leaves the condensate once it has
reached the outgoing lead. This implies that the electron spends just a finite
period of time in the condensate and therefore kicks just a finite number of
atoms out of the latter. Technically, this manifests itself in the divergency
of the integral for $P_{z}^{\prime}(\zeta,\eta,0)$ emerging from
Eq.(\ref{GenZ}) for the case $\gamma=0$. By direct integration over
$\mathrm{d}Z\mathrm{d}w$ in Eq.(\ref{dNZ}) for the case of large leads one
finds
\begin{equation}
P_{z}^{\prime}(\zeta,\eta,0)=-\frac{G_{A}G_{R}}{\pi^{2}\left(  2\mathcal{N}%
q_{s}\zeta\right)  ^{3}}. \label{NuF}%
\end{equation}

\begin{figure}
[ptb]
\begin{center}
\includegraphics[
height=2.0401in,
width=2.6567in
]%
{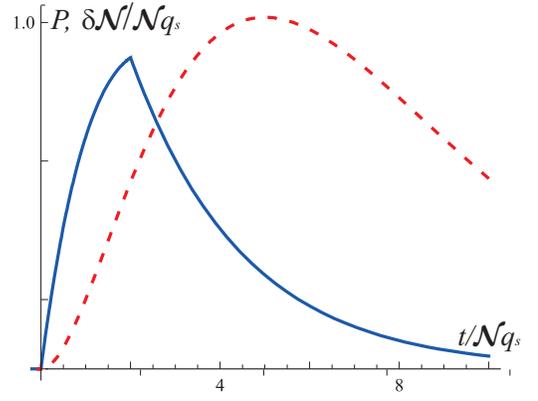}%
\caption{ Tunneling pronability $P$ (solid line) and the number of particles
$\delta\mathcal{N}$ that have left the condensate in function of time
line.\ (dashed line, not to scale) for $\gamma\neq0$.}%
\label{NdT}%
\end{center}
\end{figure}

Fourier transformation of Eqs.(\ref{TunProbS})\ref{NuF} over $\zeta$ has the
form%
\begin{align}
&  P(t,\eta,0)=G_{A}G_{R}~\frac{\left(  t\Theta\left(  t\right)  -\left(
t-2\mathcal{N}q_{s}\right)  \Theta\left(  t-2\mathcal{N}q_{s}\right)  \right)
}{2\mathcal{N}^{2}q_{s}^{2}}\label{LinTProd}\\
&  P_{z}^{\prime}(t,\eta,0)=~-\frac{q_{s}G_{A}G_{R}}{\left(  \mathcal{N}%
q_{s}\right)  ^{3}}t^{2}\Theta\left(  t\right)  , \label{DNLINN}%
\end{align}
where $\Theta\left(  x\right)  $ is the step function. Note that the time
dependence of the tunneling probability comes from the spectrum statistics; it
is typical of the chosen Gaussian unitary ensemble (GUE). In Fig.\ref{NdT} we
present these two dependences for $\gamma\neq0$,%
The finite $\gamma$ results in the net tunneling probability
\begin{equation}
\int\gamma P(t,\eta,0)\mathrm{d}t=G_{A}G_{R}\frac{1-\mathrm{e}^{-2\mathcal{N}%
\gamma q_{s}}}{2\mathcal{N}q_{s}} \label{TTP}%
\end{equation}
and the total number of deplited atoms
\begin{equation}
\int P_{z}^{\prime}(t,\eta,0)\mathrm{d}t\mathrm{d}\mathcal{V}=\frac{q_{s}%
}{4\left(  \mathcal{N}\gamma q_{s}\right)  ^{3}}\int G_{A}G_{R}\mathrm{d}%
\mathcal{V} \label{TKA}%
\end{equation}
It corresponds to the situation where the destructive interference increases
in the course of time and stops the tunneling process after a time of the
order at the Heisenberg time. At longer times, the electron does not present
in the condensate and hence does not result in atomic losses.

We now concentrate on the position of the saddle point. To this end one has to
consider the saddle-point conditions suggested by the requirement%
\[
\partial_{q_{s}}\left(  \mathcal{N}q_{s}^{2}~+\int\limits^{p_{c}}%
\frac{\mathcal{V}\mathrm{d}p^{\prime}}{\left(  2\pi\right)  ^{d}}\mathrm{\ln
}{\Huge [}\eta-p^{\prime2}\mathbf{-}q_{s}{\Huge ]}\right)  =0,
\]
where the cut-off momentum is given by the condition $\frac{\mathcal{V}%
}{\left(  2\pi\right)  ^{d}}\int^{p_{c}}\mathrm{d}^{d}p=\mathcal{N}$. One can
consider the dimensionless density $n=\mathcal{N}/\mathcal{V}$ as a small
parameter. Taking the derivative one finds
\[
2nq_{s}~=\int\limits^{p_{c}}\frac{\mathrm{d}p^{\prime}}{\left(  2\pi\right)
^{d}\left(  \eta-p^{\prime2}\mathbf{-}q_{s}\right)  }.
\]
The left hand side of the equation contains the small parameter, and therefore
the right hand side also should be small. The contribution to the integral on
the right hand side comes from the domain of small momenta $p^{\prime}\sim
n^{1/d}$ while $\eta\sim q_{s}\sim1$. We therefore can neglect $p^{\prime2}$
in the denominator and arrive at%
\begin{equation}
2nq_{s}~=\frac{1}{\left(  \eta\mathbf{-}q_{s}\right)  }\int\limits^{p_{c}%
}\frac{\mathrm{d}p^{\prime}}{\left(  2\pi\right)  ^{d}}=\frac{n}{\left(
\eta\mathbf{-}q_{s}\right)  }, \label{SPE}%
\end{equation}
which immediately results in%
\begin{equation}
q_{s}=\frac{\eta\pm\sqrt{\eta^{2}-2}}{2} \label{WL}%
\end{equation}
that is in the dependence typical of the random matrices and resulting in the
Wigner semicircle law for the state density. Imaginary part of this quantity
is the parameter $\mathrm{i}q_{s}$ entering Eqs.(\ref{GRan}), while the real
part (which have been earlier ignored in the expressions for $P$ and
$\delta\mathcal{N}$) just gives a shift of the energy scale. Substitution of
the result Eq.(\ref{WL}) to Eqs.(\ref{GRan}) yields%
\begin{equation}%
\begin{array}
[c]{c}%
G_{A;R}=\frac{n2^{-1/2}}{\sqrt{\eta\pm\sqrt{\eta^{2}-2}}}\exp\left[
\pm\mathrm{i}R\left(  \frac{-\eta\pm\sqrt{\eta^{2}-2}}{2}\right)  ^{\frac
{1}{2}}\right]  \mathrm{~~for~1D}\\
G_{A;R}=\frac{n}{2\pi}K_{0}\left(  \pm R\left(  \frac{-\eta\pm\sqrt{\eta
^{2}-2}}{2}\right)  ^{\frac{1}{2}}\right)  ~~~~~~~~~~~~~~\mathrm{for~2D}\\
G_{A;R}\rightarrow\frac{n}{4\pi R}~\exp\left[  \pm\mathrm{i}R\left(
\frac{-\eta\pm\sqrt{\eta^{2}-2}}{2}\right)  ^{\frac{1}{2}}\right]
~~~~~~~~~~~~\mathrm{for~3D.}%
\end{array}
\label{GRANF}%
\end{equation}

\begin{figure}
[ptb]
\begin{center}
\includegraphics[
height=2.1698in,
width=3.3269in
]%
{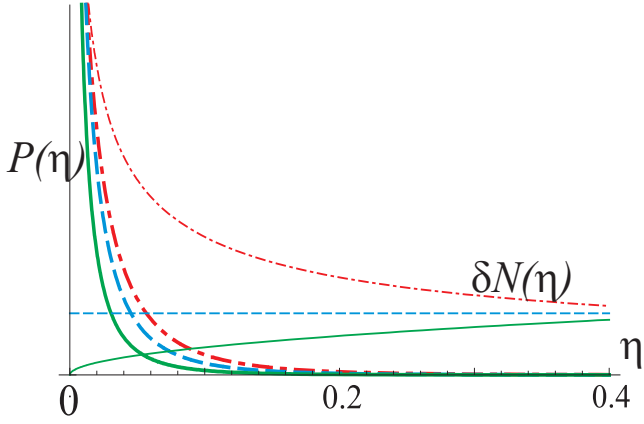}%
\caption{ Tunneling probability (bold lines) and the number of atoms kicked
out of the condensate (thin lines) as \ functions of the energy $\eta$ (not to
scale). Solid lines correspond to $1D$ , dashed lines to $2D$, and dash-dot
lines to $3D$. Zero energy corresponds to the bottom of the band $\eta
=-\sqrt{2}$. Here $R=10$ and $\gamma=0.1$.}%
\label{fpdn}%
\end{center}
\end{figure}

Now one has to substitute the obtained Green's functions to the expression
Eqs.(\ref{TTP}\ref{TKA}) for the net tunneling probability and the total
number of atoms kicked out from the condensate.
\begin{equation}
\int\gamma P(t,\eta,0)\mathrm{d}t=G_{A}G_{R}\frac{1-e^{-2\mathcal{N}\gamma
q_{s}}}{2\mathcal{N}q_{s}}%
\end{equation}
and the total number of deplited atoms
\begin{equation}
\int P_{z}^{\prime}(t,\eta,0)\mathrm{d}t\mathrm{d}\mathcal{V}=\frac{q_{s}%
}{4\left(  \mathcal{N}\gamma q_{s}\right)  ^{3}}\int G_{A}G_{R}\mathrm{d}%
\mathcal{V}%
\end{equation}
These results are shown in Fig.\ref{fpdn}%
One sees that the dependences $P(\eta)$ are not too much sensitive to the
dimensionality of the problem, and the main contributionto the conductance
comes from the edge of the electron levels band, which is formed as the result
of the perturbation of the mean-field band by the strong quantum fluctuations
of the atomic density in the condensate. The dependences $\delta
\mathcal{N}(\eta)$ have more pronounced dependence on energy via dependence of
the volume ocupied by the tunneling electron on the spacial dimensionality.


\begin{thebibliography}{99}
\bibitem {IZ}That is, involving only a single $Q$ configuration stem from the
semiclassical trace or Itzykson--Zuber integral formulae.
\end{thebibliography}
\end{document}